\documentclass[preprint,aps,tightenlines,floatfix,showpacs]{revtex4}

\usepackage{graphics}
\usepackage{bm}
\usepackage{amsmath}
\usepackage{amssymb}

\begin{document}

\title{Comments on methods of analyses of nucleon-nucleus scattering data}

\author{K. Amos$^{1}$} 
\email{amos@physics.unimelb.edu.au}
\affiliation{School of Physics, University of Melbourne,
Victoria 3010, Australia}

\author{S. Karataglidis$^{2}$}        
\email{stevenka@ac.uj.za}
\affiliation{Department   of  Physics,  University   of  Johannesburg,
  P.O. Box 524, Auckland Park, 2006, South Africa}

\date{\today}

\begin{abstract}
  Methods  of   analysis  of  nucleon-nucleus   scattering  data  have
  progressed markedly  over the  past 50 years,  yet many  analyses of
  scattering data still use  prescriptions specified at various stages
  of that  progress.  But  the assumptions made  or inherent  in those
  analyses are  poorly justified and  most are no longer  necessary to
  facilitate  evaluations.  We investigate  some of  those assumptions
  and highlight  deficiencies and the  uncertainties they may  make in
  data analyses.
\end{abstract} 

\maketitle


\section{Introduction}

An ultimate  aim of a nuclear  reaction theory is to  probe details of
the  structure of  the nuclei  involved.  There  are many  and diverse
reactions and  observables one might use  in that quest  but herein we
consider  just cases  of nucleon  scattering (elastic  and inelastic),
focussing  upon questions  that need  be raised  regarding  methods of
analyses  used.   The questions  raised  apply  to  various extent  to
theories  for the  scattering  and reactions  involving other  nuclear
projectiles.

First we deal with the  problem of elastic scattering of nucleons.  It
is of great  regret that many assume that  elastic scattering data are
not  of  interest experimentally  or  theoretically.   But almost  all
theories  of  nucleon-nucleus reactions  presume  a  knowledge of  the
elastic scattering  data for the  system. It is commonly  thought that
such can be approximated by using some global, phenomenological, local
optical  model potential.  But  in doing  so, there  are uncertainties
that may vitiate using a  reaction theory for data analysis to achieve
the prime purpose of assessing the structure of the target.

For many years analyses  of experimental data from reactions initiated
by nucleons, by protons mostly, have  been of interest, in part as the
incident proton  most strongly interacts with neutrons  in the target.
Analyses of  proton scattering data then complement  those of electron
scattering  form  factors  which  are  more sensitive  to  the  proton
distributions   in  nuclei.    If,   as  well,   static  moments   and
electromagnetic  transition rates are  known, the  body of  data could
well  assess putative  details  of low  energy  structures of  nuclei.
However,  the methods  of data  analysis  used need  be of  comparable
surety for any conclusions  drawn to have sufficient credibility.  But
one must bear in  mind that electromagnetic transition rates, electron
scattering form factors, and  nucleon scattering data can be sensitive
to disparate aspects of the nuclear  wave functions so that a model of
structure may be adequate for  some data analyses and not others.  Now
while  the theories  used  to analyze  electromagnetic  data are  well
established, with the size,  if not specific influence, of corrections
to results found  known, that is not the case  with most theories used
to analyze nucleon-induced reactions.

To date  analyses of data from  nucleon induced reactions  can only be
made  using  significant approximations  to  a  basic  theory of  such
scattering. In  doing so, guidance  from some physical  properties and
physics principles are needed to  justify whatever model theory of the
reaction  is used.  It  is an  evolutionary process  and one  can only
assess approximations  used in the  past against results  from another
developed theory  in which such  approximations have not been  made or
are of lesser influence.  To  date, of course, all applicable theories
of nuclear scattering involve  approximations.  Still, by treating the
system more assiduously, it  is possible to understand why conclusions
drawn from past  analyses need be treated cautiously  and not taken as
definitively as authors may have indicated.

For  analyses of  nucleon-nucleus elastic  scattering data,  the first
theoretical   approach   was    phenomenological,   dating   back   to
Bethe~\cite{Be40}. The  form chosen  for the associated  optical model
potentials was  taken to reflect the expected  density distribution of
nucleons in each target  nucleus.  Subsequently, optical potential was
sought  microscopically, folding  a  free nucleon-nucleon  interaction
with a  density profile for the  target. The most well  known of those
approaches was that of  Kerman, McManus, and Thaler (KMT)~\cite{Ke59}.
Later  formulations  took  into  account  medium  corrections  to  the
interaction  between the  incident  nucleon with  those  bound in  the
target as well as  specifically including the nonlocality arising from
exchange amplitudes associated with  treatment of the Pauli principle.
Those   $g$-folding   models   are    discussed   in   detail   in   a
review~\cite{Am00}.
 
All  of the  models identified  in the  above, and  various  sets with
intermediate steps  (such as the JLM  approach~\cite{Je76}) have been,
and continue to  be, used to analyse scattering  data. What we present
is not  intended as a  justification of the most  recent applications.
Rather we seek  to point out the inadequacies of  the models formed in
distinct stages of development, why some should no longer be used, and
why the  conclusions drawn from the  use of many be  treated with more
caution.

\section{The optical model potentials}

Most  optical potentials  used  are  of local  type,  whether they  be
defined  phenomenologically  (usually  with  Woods-Saxon form)  or  by
folding schemes. Typical  of the latter is the  JLM method as recently
used~\cite{Be06}, in which the direct (local) term is found by folding
an $NN$  interaction with the  nucleon density profile of  the target.
The imaginary term is specified phenomenologically.  Some also account
approximately   for   the    exchange   process   arising   from   the
indistinguishability of the projectile  with the nucleons bound in the
target by using an additional  local term. But there are problems with
all of these  approaches, and especially in using  the relative motion
wave functions generated from local optical potentials.

\subsection{On fitting elastic scattering data}

A primary problem,  and a rather insidious one,  is that justification
of  an optical  potential is  founded upon  finding a  fit  to elastic
scattering data:  cross sections and  spin observables.  Of  course, a
viable  optical  potential  in  a  single channel  theory  of  elastic
scattering should  indeed provide such fits.   But does a  fit to such
data imply a proper  (the ``correct'') optical potential?  Many assume
that  the  answer  to the  question  is  yes.   Few, it  seems,  fully
appreciate the limited validity of such a justification.

A fit  to elastic  scattering data only  requires that  the scattering
theory produce a  set of phase shifts that, when  used in the standard
sums of Legendre  polynomials, give a good fit to data.   In no way is
that a proof  of the uniqueness of the phase shifts,  and even less of
the interactions  used in the Schr\"odinger  equations producing them.
Of course, a  ``proper'' optical potential must lead to  a good set of
phase shifts, and  as well a fit to  measured elastic scattering data,
but in specifying the interaction,  one must use as many principles of
physics as  possible to have maximal credibility.   Basically, all the
usual phenomenological and part  phenomenological models that end with
purely  local forms for  a nucleon-nucleus  optical potential  fail in
that respect  to some extent.  To get elastic scattering  phase shifts
only  requires knowledge  of the  relative motion  wave  functions and
their derivatives  at extreme separation  of the nucleon  and nucleus.
How those  wave functions attained the  required asymptotic properties
from  solutions of  Schr\"odinger equations  is not  defined  by those
numbers alone.

Given a  set of phase shifts  that typically are necessary  to have an
excellent fit to data, such as  with the elastic scattering of 200 MeV
protons  from $^{12}$C,  phenomenological potentials,  potentials from
inverse   scattering  theories,   and  those   microscopically  formed
($g$-folding) all exist from which  quality fits to the same data have
been found.  Some of those  fits are associated with a chi-squared per
degree of freedom ($\chi^2$/F) values $\sim 1$.

\subsubsection{An example of equivalent optical potentials}

\begin{figure}
\scalebox{0.6}{\includegraphics*{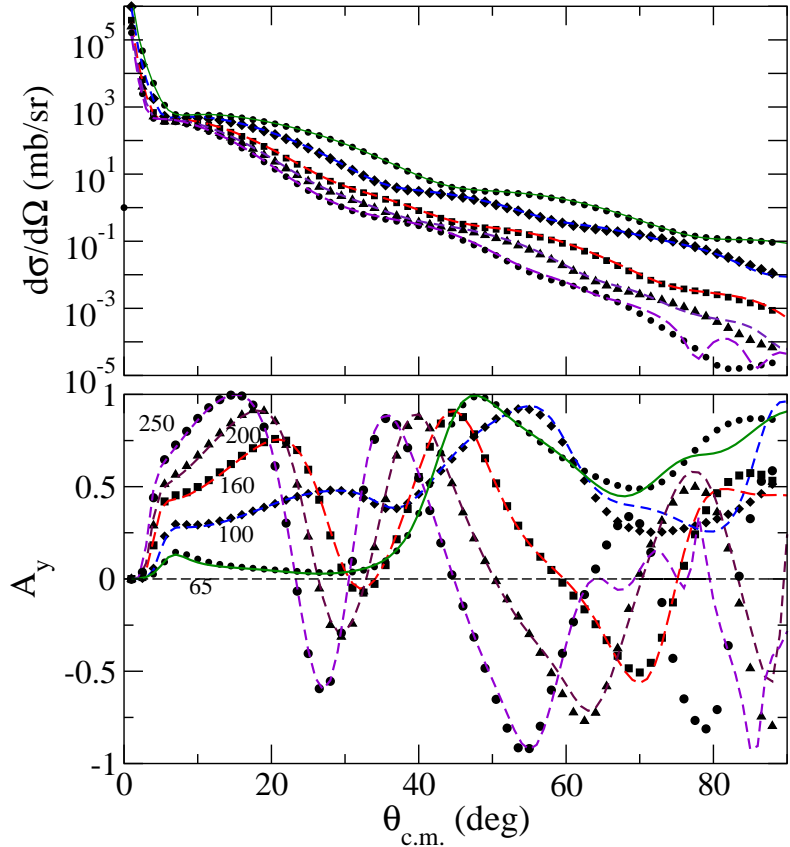}}
\caption{\label{p-12C-60-250}(Color   online)   Comparisons   of   the
  differential cross sections (top) and analyzing powers (bottom) from
  defined  sets of  phase shifts  with  those deduced  by using  their
  inversion potentials.}
\end{figure}

The   conventional  approach,   the  phenomenological   optical  model
potential  approach, to  analyses  of elastic  scattering  data is  an
example  of inverse  scattering.   Specifically it  is  an example  of
numerical inverse scattering: the fit to data determines values of the
parameters  deemed  best.   More  generally,  the  inverse  scattering
problem  for  fixed  energy  scattering  resolves  to  the  following:
\textit{Given an  $S$-function (phase shifts) at  a particular energy,
  find  local central  and spin-orbit  potentials that  reproduce that
  $S$-function  when  used  in  Schr\"odinger equations.}   One  fully
quantal   method  to   do   that  is   the  modified   Newton-Sabatier
scheme~\cite{Lu98} which finds potentials of the Bargmann class.  That
approach  has been  used  with  sets of  phase  shifts (the  ``data'')
obtained  from  a  microscopic  ($g$-folding)  model  of  the  elastic
scattering of  protons at energies  of 65, 100,  160, 200 and  250 MeV
from  $^{12}$C.   The  cross   sections  and  spin  measurables  found
therefrom~\cite{Am00} are in very  good agreement with actual measured
data.  For comparison,  those phase shifts have been  used as input to
the   inversion   scheme~\cite{Lu98}   to   find  a   set   of   local
(Schr\"odinger)  potentials~\cite{Lo00}.   The  (complex, central  and
spin-orbit) potentials  were then  used in Schr\"odinger  equations to
find if the  same set of phase shifts would  be regenerated. The cross
sections and analyzing powers from  the $g$-folding model are shown by
the symbols in Fig.~\ref{p-12C-60-250} while those resulting from
use of  the determined (local)  inversion potentials are shown  by the
various dashed  lines.  Clearly the  reproductions are very  good with
but  a minor mismatch  for 250  MeV at  large ($\theta  \ge 70^\circ$)
scattering angles.

\begin{figure}
\scalebox{0.6}{\includegraphics*{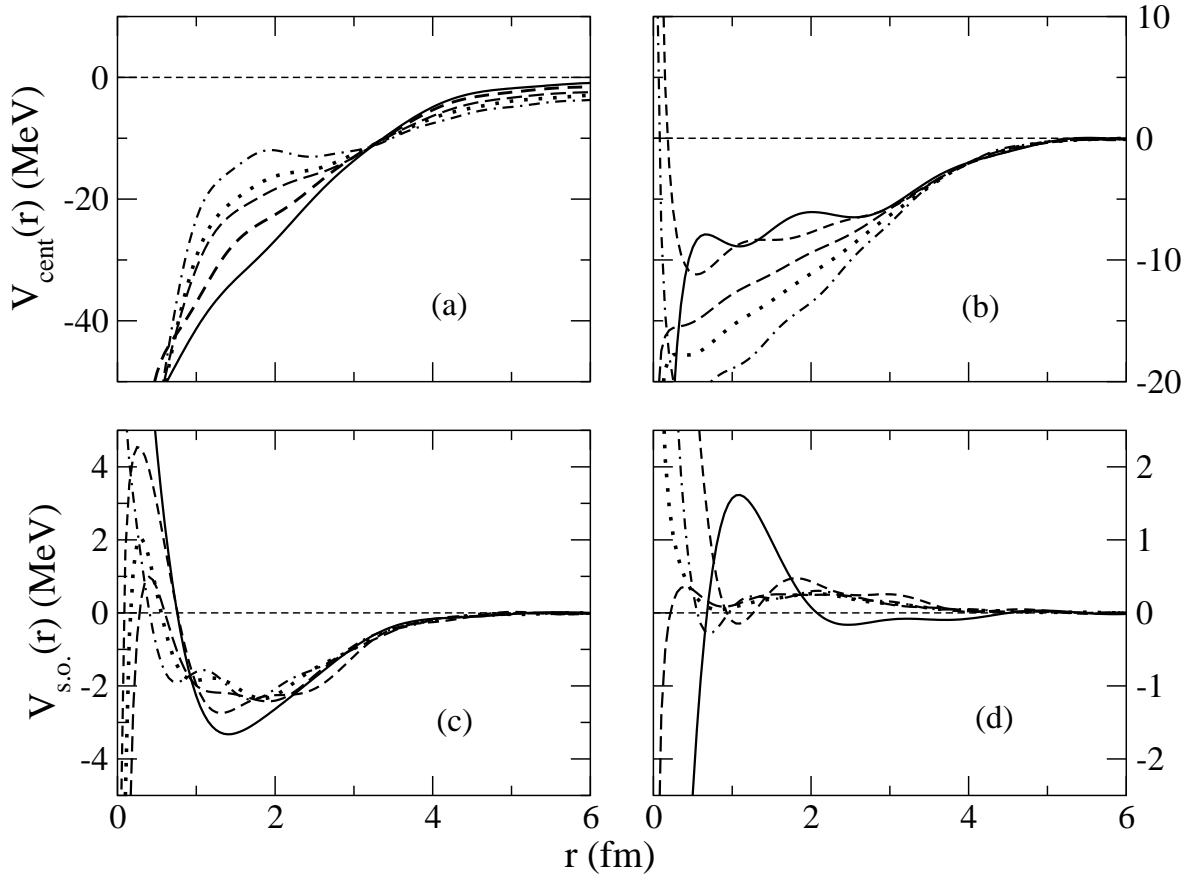}}
\caption{\label{pCinvpots} The potentials  defined by using a modified
  Newton-Sabatier inversion scheme with the set of phase shifts for 65
  to 250 MeV protons on $^{12}$C.}
\end{figure}

The  potentials  obtained   from  modified  Newton-Sabatier  inversion
calculations are displayed in Fig.~\ref{pCinvpots}.  Therein the
central  real,  central  imaginary,  spin-orbit real,  and  spin-orbit
imaginary terms  are displayed in  panels labelled (a), (b),  (c), and
(d) respectively. The  potentials for energies 65, 100,  160, 200, and
250 MeV are  portrayed by the solid, dashed,  long-dashed, dotted, and
dot-dashed curves  respectively. The energy  dependence resulting with
these interactions  clearly is severe.  The  central potentials become
less  refractive and  more absorptive  with increase  in  energy.  The
spin-orbit  interactions   vary  even  more   markedly.   These  local
potential forms  determined by  quantal inverse scattering  theory are
distinctly different  from the Woods-Saxon  interactions usually taken
with  the phenomenological optical.   But they  fit the  ``data'' (and
from  the origins,  the actual  measured data)  very well.   Indeed, a
criterion was that the inversion  process must give ``data'' fits with
measure $\chi^2$/degree of freedom of $\mathcal{O}(1)$.

\begin{figure}
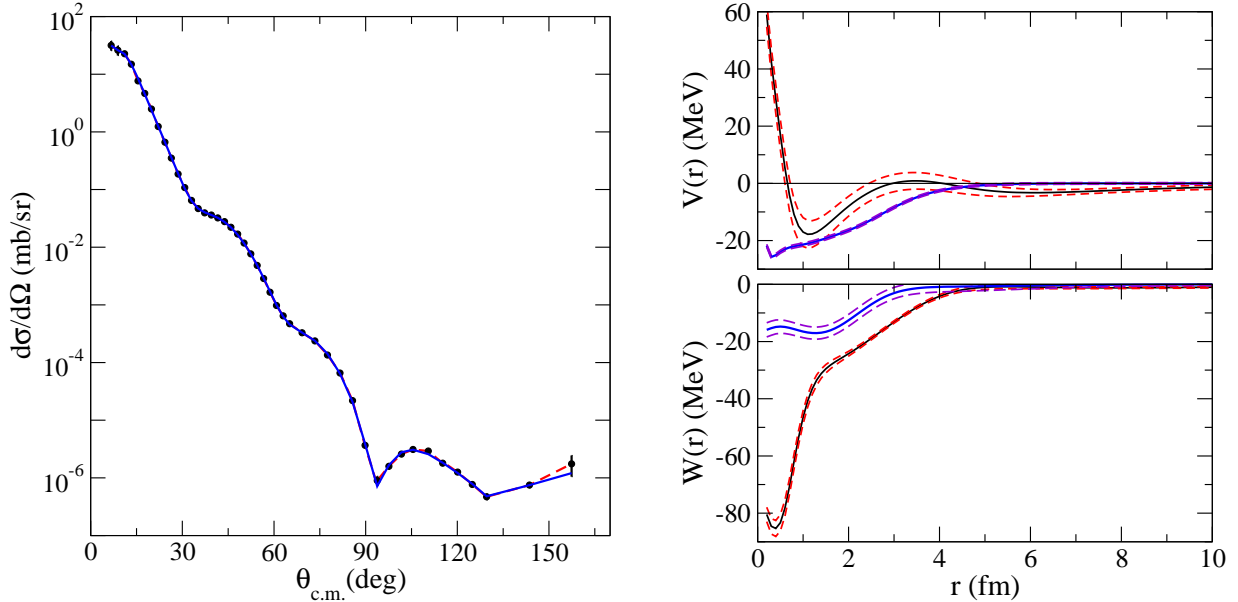

\begin{minipage}[b]{0.45\linewidth}
\scalebox{0.47}{\includegraphics*{Figure3a.eps}}
\end{minipage}\hfill
\begin{minipage}[b]{0.47\linewidth}
\scalebox{0.48}{\includegraphics*{Figure3b.eps}}
\end{minipage}\hfill
\caption{\label{pC200}(Color  online)   The  data  from   the  elastic
  scattering  of  200 MeV  protons  on  $^{12}$C  compared with  cross
  sections determined from two  inversion potentials (left panel). The
  two potentials  are depicted,  as functions of  radius in  the right
  panel. The real and imaginary parts  are shown in the top and bottom
  segments    respectively   with    their    bands   of    confidence
  indicated. Details are given in the text.}
\end{figure}

The actual  cross section data~\cite{Me83} for  the elastic scattering
of 200 MeV  protons on $^{12}$C are numerous and  of high quality. The
basic  Newton-Sabatier method has  been used~\cite{Al94}  to ascertain
local complex potentials from them.  Two distinct inversion potentials
have   been  found.    They  are   shown   in  the   right  panel   of
Fig.~\ref{pC200}; the real parts in  the top and the imaginary ones in
the bottom segments.  The loop  was completed by using both potentials
in Schr\"odinger equations whose solutions gave phase shifts, and then
the scattering cross  sections that are compared with  the actual data
in the  left panel of Fig.~\ref{pC200}.  The  evaluated cross sections
are practically  indistinguishable from  the data reflecting  that the
two results  fit the  data with values  of $\chi^2$/degree  of freedom
1.006 and 1.018.  The two calculated cross sections differ only in the
region of the largest scattering angle, and then but slightly.

The two distinctive  potentials were found from two  sets of pole-zero
pairs that  represent the  $S$-function that one  needs first  for the
chosen   inverse   scattering    theory.    Details   are   given   in
Ref.~\cite{Al94}.  The approach allows specification of bands of
confidence on those potentials.  They are shown in Fig.~\ref{pC200} by
the dashed curves surrounding  the potentials. Those bands, determined
from an error analysis~\cite{Al94},  are to be interpreted as follows:
should another potential exist associated  with a fit to the data with
the same  $\chi^2$/F, then there  is a 60\%  chance that it  would lie
within  that  band.  Clearly  then,  as the  two  cases  shown are  so
different,  even  such  good  data  are insufficient  to  delineate  a
potential  uniquely.   To  these   one  might  add  any  finely  tuned
phenomenological  model form.   Thus  a claim  that a  nucleon-nucleus
interaction is the  correct one, based solely upon  its fit to elastic
scattering data, is not valid.   A ``correct'' potential must be based
upon more  physical information.  Simply linking its  form to expected
matter and  charge distributions of  the nucleus is not  sufficient in
this regard, however.

\subsection{The effect of the Pauli principle}

A problem with most single  channel (optical model) potentials is that
they  are   local  in  form.    Those  defined  fully,  or   in  part,
phenomenologically usually take parametric  forms and values in accord
with some notions regarding  the matter and/or charge distributions of
nucleons in the  targets.  The charge distribution of  the targets may
be well  reflected by the form  factor from the  elastic scattering of
electrons but one  cannot assume that the matter  distribution is then
necessarily as  well-described.  This is most  evident when describing
scattering from or by  neutron halo nuclei~\cite{St02}.  Certainly the
elastic  scattering of  nucleons does  not directly  justify parameter
forms for the reasons discussed above.

\begin{figure}
\scalebox{0.6}{\includegraphics*{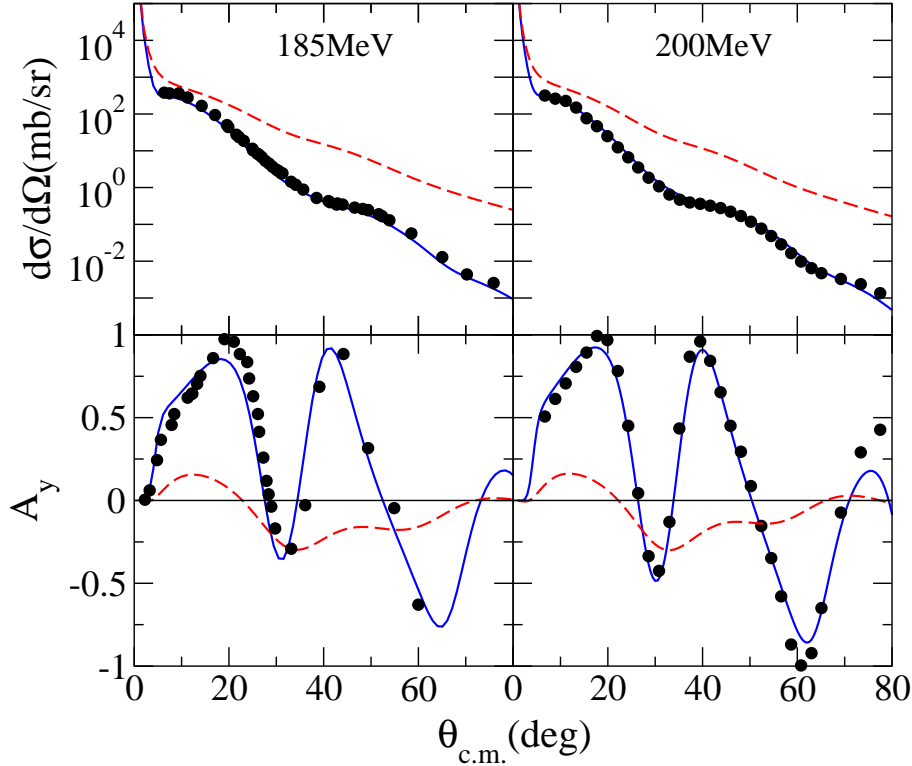}}
\caption{\label{nonloc}(Color  online)   Differential  cross  sections
  (top) and  analyzing powers (bottom) from the  elastic scattering of
  185 (left)  and 200  MeV (right) protons  with $^{12}$C.   The solid
  curves  are the  results found  using a  complete  $g$-folding model
  optical potential  while the dashed curves are  the results obtained
  by omitting the nonlocal exchange parts of those interactions.}
\end{figure}

One may hold  faith in the neutron distributions  given by large basis
models of  the ground state structure  of any nucleus  at least inside
and  near  to the  nuclear  surface.   Indeed,  their use  in  folding
$NN$-interactions to form an $NA$ optical potential lead to results in
good agreement with elastic scattering observables.  But, in so doing,
unlike  the case  of  electron  scattering, one  must  also take  into
account  effects of  the Pauli  Principle.  The  prime effect  of that
principle  in  defining $NA$  optical  potentials  in the  microscopic
approach built  essentially on KMT theory~\cite{Ke59}, is  to make the
interaction highly  nonlocal. That nonlocality is  associated with the
exchange  amplitudes  arising  from  the indistinguishability  of  the
projectile  with  nucleons  in  the target.   The  contributions  from
scattering  in which  the detected  (emergent) nucleon  originally was
bound within the target are  extremely important. They must be treated
carefully, since the associated (exchange) amplitudes have a different
momentum transfer  profile to the  direct scattering ones.   Even more
critical  is  that,  often,  these exchange  amplitudes  destructively
interfere  with the  direct ones.   This has  been demonstrated  for a
range  of  energies~\cite{De00} so  only  an  illustrative example  is
included herein. Cross sections from the elastic scattering of protons
from  $^{12}$C  at   energies  of  185  and  200   MeV  are  shown  in
Fig.~\ref{nonloc}.
Masking  of   such  effects  in   the  fitting  procedures   of  local
phenomenological forms is  insufficient. As discussed previously, such
potentials are necessarily not unique.

The considerable effects of the exchange amplitudes indicate also that
theories predicated upon just  the charge and/or matter densities have
a  problem. The  exchange  amplitudes require  input  of the  one-body
density matrices which include, but contain more information than, the
simple density.   One needs explicit single nucleon  wave functions in
fact.   The  severe destructive  interference  noted  also means  that
seeking equivalent  local potentials to approximate such  may be quite
problematic.

\subsection{Medium effects with the basic $NN$ interaction}

In the  1980's there was a watershed  in the studies of  $NN$ and $NA$
scattering.  First,  reliable $NN$ scattering  amplitudes were defined
and  detailed  phase shift  analyses~\cite{Ar83}  made,  at least  for
energies up  to the pion threshold.   Second the Nijmegen~\cite{Na78},
Paris~\cite{La80}, and Bonn~\cite{Ma87} $NN$ potentials were developed
to fit  those $NN$  amplitudes and phase  shifts.  Also the  status of
microscopic $NA$ optical potential  theories was reviewed at a seminal
topical workshop in Hamburg~\cite{Ge79}. Finally, experimental studies
in those  years produced many and  varied high quality  $NN$ data sets
for energies  to 1~GeV~\cite{Sc83}, adding  impetus to implementations
of  the theories.   Since  that time  reviews~\cite{Ra92,Am00} of  the
optical model for scattering have  been made given the central role it
plays in studies of nucleon-induced reactions~\cite{Sa83,Fe92,Ho94}.

Standard  now is  the  optical potential  built upon  non-relativistic
multiple  scattering  theory   with  the  $NN$  scattering  amplitudes
modified   from   the   free   $NN$   values.   The   result   is   an
\textit{effective $NN$ interaction}. Those modifications are caused by
the two nucleons interacting within  the nuclear medium and are due to
\textit{Pauli  blocking}  and  \textit{mean  field} effects  for  both
projectile  and bound state  nucleons.  In  addition, there  are other
effects due to the convolution  of the $NN$ scattering amplitudes with
target   structure   that   require  off-the-energy-shell   scattering
amplitudes.  Also of importance is the complete antisymmetrization (at
least at the two nucleon level) of the $A+1$ nucleon scattering system
which  leads to direct  and knock-out  (exchange) amplitudes  for $NA$
scattering.  As  illustrated before, the effect of  such $NA$ exchange
amplitudes is not small at any energy and those amplitudes are a major
source of nonlocality.  Other medium effects~\cite{Ra90} also can have
some significance.  Details of  the forming the effective interactions
from   a   chosen   free   $NN$   interaction   are   given   in   the
review~\cite{Am00}.   We  have  used  both  the Paris  and  the  BonnB
interactions  in  our  evaluations   of  $NN$  $t$-matrices  and  $NN$
$g$-matrices; the choice gives only small differences in results.

\begin{figure}
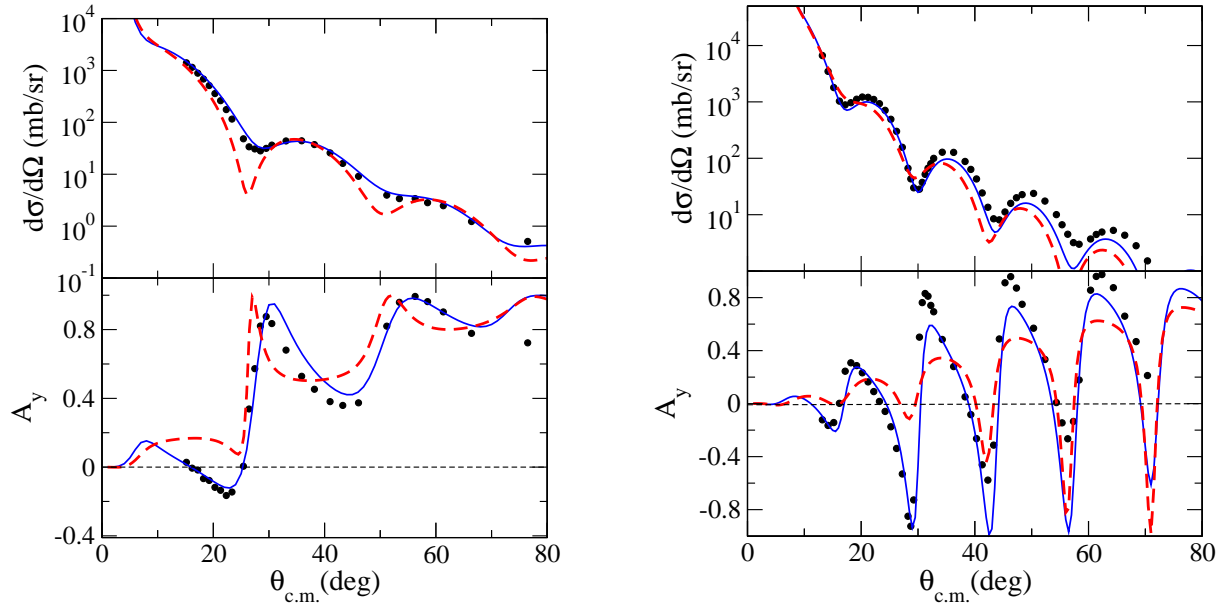

\begin{minipage}[b]{0.47\linewidth}
\scalebox{0.47}{\includegraphics*{Figure5a.eps}}
\end{minipage}\hfill
\begin{minipage}[b]{0.48\linewidth}
\scalebox{0.48}{\includegraphics*{Figure5b.eps}}
\end{minipage}\hfill
\caption{\label{dd-free}(Color   online)   Elastic  scattering   cross
  sections  and analyzing powers  for 65  MeV protons  scattering from
  $^{40}$Ca (left) and from  $^{208}$Pb (right).  The solid curves are
  the  results of using  the density  dependent, $NN$  $g$-matrices in
  forming  the optical  potentials, while  the dashed  curves  are the
  results  found  using the  free  $NN$  $t$-matrices  in the  folding
  process.}
\end{figure}

In Fig.~\ref{dd-free} the effects that medium dependence of the $NN$ 
effective interaction have in  the cross sections and analyzing powers
are shown.  Results for the  elastic scattering of 65 MeV protons from
$^{40}$C are shown  in the left panel while  those from $^{208}$Pb are
given  in the  right panel.   The  data~\cite{Sa82} are  shown by  the
filled circles  while the  calculated results found  using $g$-folding
potentials are depicted by the  solid curves. The results shown by the
dashed curves are from optical potentials having no density dependence
in the  effective interaction, i.e.   formed by folding the  free $NN$
$t$-matrices. The  density matrices for $^{40}$Ca used  in forming the
folding   potentials  were   taken   from  SHF   (Skyrme-Hartree-Fock)
calculations~\cite{Ka02} while those for $^{208}$Pb were from a simple
packed  shell model  with  oscillator wave  functions ($\hbar\omega  =
6.7$~MeV).

\begin{figure}[h]
\scalebox{0.6}{\includegraphics*{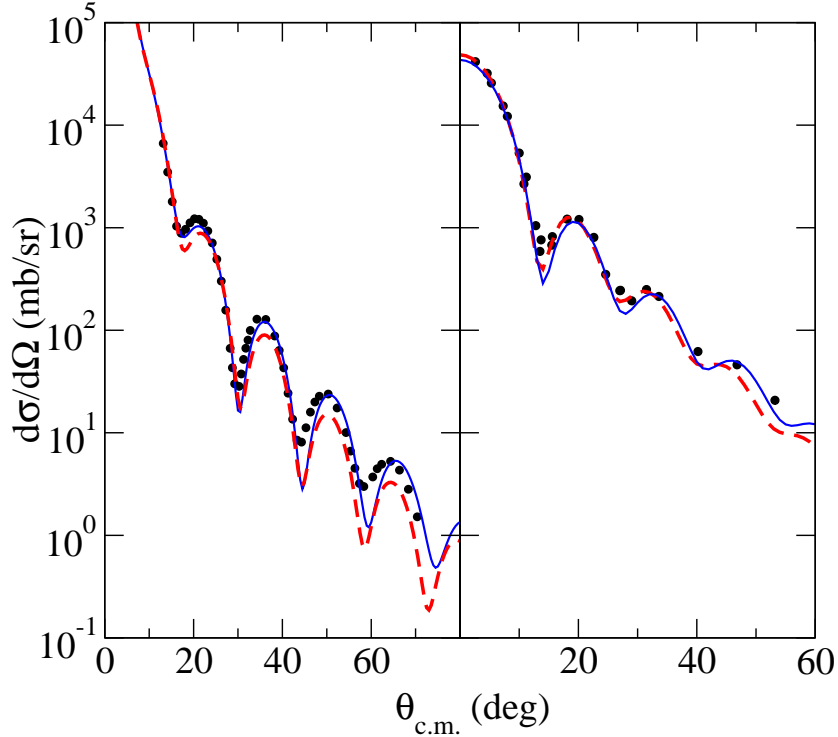}}
\caption{\label{Pb-skm}(Color online)  Cross sections for  the elastic
  scattering  of   protons  (left)   and  of  neutrons   (right)  from
  $^{208}$Pb.}
\end{figure}

The role of  density dependence in the $NN$  effective interactions is
very evident,  most notably in  the analyzing power  results.  Density
dependence  clearly improves agreement  with the  data in  both cases.
However, it is apparent that the $^{208}$Pb results are not in as good
agreement with the data as are the $^{40}$Ca ones. The issue is one of
structure as is shown next in Fig.~\ref{Pb-skm}.  In this
figure, cross sections from the  elastic scattering of 65 MeV nucleons
from    $^{208}$Pb      are    shown.     The   data    from    proton
scattering~\cite{Sa82}  and  from  neutron scattering~\cite{Ib00}  are
compared  with results  from  $g$-folding potentials  found using  two
different models of the nuclear wave functions.  The dashed curves are
results  given by  a packed  shell  model with  proton oscillators  of
$\hbar\omega = 6.7$  MeV as before, but $\hbar \omega  = 7.25$ MeV was
used for the neutrons.  This  gave a neutron skin thickness ($S= 0.16$
fm), comparable to that  of an SHF calculation~\cite{Ka02}.  The solid
curves in the figure result when  SHF model wave functions are used to
specify  the  $g$-folding potentials.   The  neutron scattering  cross
sections are quite well reproduced  by both calculated results but the
proton  scattering results  indicate a  preference for  the  SHF model
input.   This  reflects  primarily   the  difference  in  the  neutron
distributions given by the two models of structure.

\begin{figure}
\scalebox{0.6}{\includegraphics*{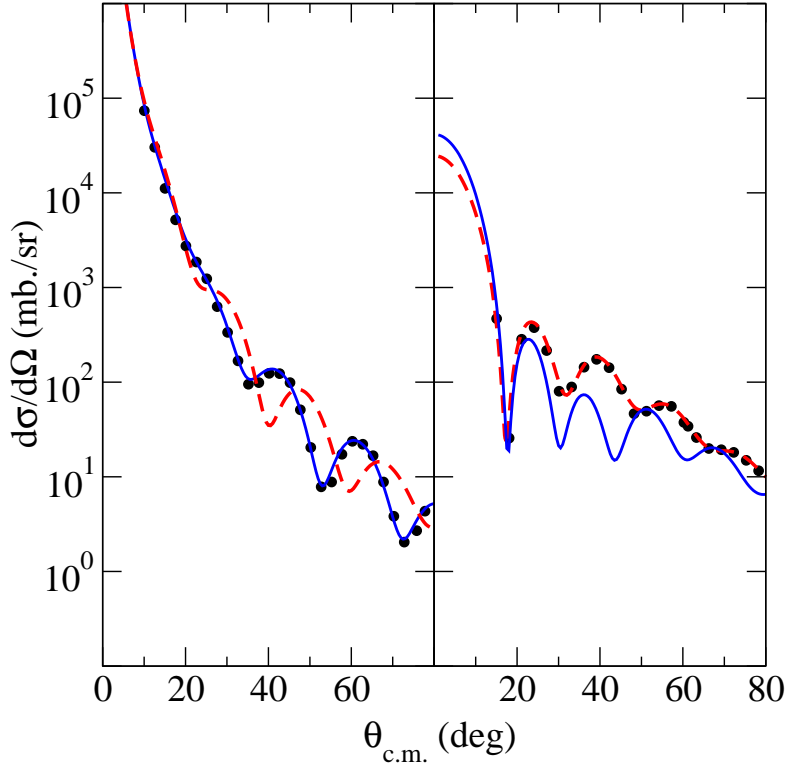}}
\caption{\label{Pb-comp}(Color online) Cross  sections for the elastic
  scattering of 40  MeV protons (left) and of  40 MeV neutrons (right)
  from  $^{208}$Pb, compared  to results  from  using phenomenological
  optical model potentials.}
\end{figure}

Of course,  by judicious  adjustment of parameter  values in  a purely
phenomenological  approach,  even better  fits  to  this  data can  be
achieved. Such is shown in Fig.~\ref{Pb-comp}.  The parameter values
used were those defined by van Oers \textit{et al.}~\cite{vO74} for 40
MeV  protons incident  on $^{208}$Pb,  and by  Finlay  \textit{et al.}
\cite{Fi84} for  40 MeV neutron  scattering.  Results found  using the
van Oers  \textit{et al.} potential for both  scatterings are depicted
by the solid curves while those found using the Finlay \textit{et al.}
potential are  displayed by the  dashed curves.  The  parameter values
differ significantly as one may  expect for nucleon scattering from $N
\ne Z$ systems.  For completeness  those parameter values are given in
Table~\ref{OMparams}.
\begin{table}
\begin{ruledtabular}
\caption{\label{OMparams} The  parameter values for  the optical model
  potentials  that lead  to the  fits to  cross-section data  shown in
  Fig.~\ref{Pb-comp}.  Other specifics  are given in Refs.~\cite{vO74}
  and \cite{Fi84}}
\begin{tabular}{lcc}
 term & \ \ 40 MeV protons &\ \ 40 MeV neutrons\\
\hline
$V_0$ (MeV) & 51.8\ \ & 36.73\  \\
$r_0$ (fm)  & 1.159 & \ 1.205 \\
$a_0$ (fm)  & 0.784 & \ 0.685 \\
\hline
$W_0$ (MeV) & 3.98\ & 5.2\ \ \\
$W_D$ (MeV) & 5.75\ & 2.55\ \\
$r_D$ (fm)  & 1.321 & 1.283 \\
$a_D$ (fm)  & 0.727 & 0.569 \\
\hline
$V_{s.o.}$ (MeV) & 6.66\ & 5.75\ \\
$r_{s.o.}$ (fm) & 1.044 & 1.105 \\
$a_{s.o.}$ (fm) & 0.905 & 0.499 \\
\hline
$r_c$ (fm) & 1.05\  & \\
\end{tabular}
\end{ruledtabular}
\end{table}

\subsection{Energy dependence problems}

The projectile  energy dictates what  approximations may be made  to a
`full'  scattering  theory.   At  low  values  of  incident  energies,
discrete state effects  are known to influence scattering~\cite{Am03}.
In those  cases, a coupled-channel  model of scattering  is essential.
An  appropriate one,  which ensures  that the  Pauli principle  is not
violated even  with a collective model prescription  for the coupling,
now exists  and has  been used to  explain compound  nucleus structure
even in  exotic, radioactive light  mass systems~\cite{Ca06}. Treating
the  Pauli principle  adequately in  coupled-channel  calculations was
crucial  as  shown  in  a  comparative study  of  two  coupled-channel
evaluations~\cite{Am05}.

Consider now,  incident energies above any excitation  energy value of
giant resonances (if such exist in the target).  Some time ago, it was
shown   that  specific   properties  (the   giant   resonances)  could
dynamically influence  proton scattering for energies  of protons that
coincide with their virtual excitation in the target~\cite{Ge75}.  For
such incident energies there  has been another coupled-channels method
proposed~\cite{To01}. That  method, the continuum  discretised coupled
channel (CDCC) method, has been  used with some success in analyses of
break-up reactions  for projectile energies  similar to those  we have
used.   Those successes might  be considered  indicative of  a problem
with the $g$-folding approach for nucleon elastic scattering. However,
the discretization of  the continuum is only a  statistical one and is
not  based upon  any attributes  of  the target  continuum.  Also  the
effects  of the  Pauli principle  (projectile-bound nucleons)  are not
taken into account with  the phenomenological optical potentials used.
But  there  are other  concerns  with  the  CDCC method.   As  Deltuva
\textit{et  al.}~\cite{De07}  noted:  \textit{``CDCC results  obtained
  with  two different  basis  sets,  namely the  basis  set using  the
  continuum of  the projectile  in the entrance  channel, and  the one
  using the  continuum of the  composite system in the  final transfer
  channel, led  to substantially different breakup  cross sections for
  $p (^{11}\text{Be},^{10}\text{Be})pn$.  These findings raise concern
  about  the accuracy  of the  CDCC  method, not  only as  a means  to
  describe reaction dynamics, but also  as an accurate tool to extract
  structure information on halo nuclei.''}

For medium energies,  such as we consider herein,  then such continuum
coupling as  well as coupling  between the low excitation  energies in
the target are not expected  to be important. Certainly there has been
no need for such when a  good model of structure and a reasonable $NN$
force were used  in  evaluations.  That  is  so at  least for  elastic
scattering cross sections usually  greater than about 0.1~mb/sr. We do
not dispute a role of coupled channels in a scattering process, but we
are  convinced that  such are  a  requirement in  analyses of  nucleon
elastic scattering  data only when there are  specific (collective and
not  too spread) states  in the  target nucleus  at excitation  in the
vicinity of  the incident energy  value.  Likewise there can  be other
reaction  processes contributing to  a given  transition, such  as the
formation of a  virtual deuteron via a $p$-$d$-$p$ chain as considered 
by some~\cite{Sk05,Ma10}.  But such second order  terms, even if it is 
possible to effect a virtual deuteron and   given the multiple  final, 
possibly unbound, states to which the break-up could lead, should give 
small contribution in comparison to the direct ones in forming elastic 
scattering cross sections.     Further the other problems as  detailed 
before with either the  modified JLM model or the CDCC coupling scheme 
that has  been used~\cite{Sk05,Kh01} make the estimates  given for the 
virtual formation  process unreliable.     These problems detract from  
conclusions  made  in  those  papers  on the  structures of the nuclei 
involved.

Finally, as  the incident nucleon  energy increases to near  and above
the Delta resonance  threshold in the $NN$ frame,  both that and other
resonances need  be taken into  account in defining an  effective $NN$
interaction. Relativity also becomes an increasing issue,   and one of  
more than a kinematic factor.       However, below about 300 MeV,  the 
non-relativistic  $g$-folding results compare  well with data and more 
than favorably with global Dirac potential model results~\cite{De05}.

\section{Inelastic scattering of nucleons}

Inelastic nucleon  scattering data from excitation  of discrete states
in  nuclei usually  has  been  analysed using  a  DWA (distorted  wave
approximation).  Coupled  channels analyses  also have been  made, but
they still are made using collective models of structure, usually with
local  coupled channel potentials.  That localisation suffers from the 
same    problems    as   does    the    local   phenomenological    or
semi-phenomenological optical model potentials described previously.

For elastic  scattering one only requires  appropriate relative motion
wave functions at large distances; appropriate meaning that they yield
a set of suitable phase shifts.  For non-elastic reaction evaluations,
with the  DWA, the  relative motion wave  functions need  be specified
through the nuclear volume.  The  choice of optical model form then is
of serious concern.

\subsection{Relative motion wave functions}

The problem  of not considering  the effects outlined in  the previous
sections (Pauli blocking, mean-field effects, and nonlocality) extends
to  descriptions of  nucleon-induced reactions  for which  the optical
potential  serves to  define the  effective interaction  mediating the
reaction.  In  that  respect,  an appropriate  (physically  justified)
determination of  the relative motion wave  function (nucleon incident
on  the  nucleus) is  required  in order  to  have  confidence in  the
description of the related  reaction.  In this subsection, we consider
the relative  motion wave functions associated with  two optical model
potentials.  Specifically  we   compare  the  partial  wave  functions
generated  from  a $g$-folding  model  \cite{Am03}  and  those from  a
phenomenological  potential whose  parameters have  been set  to `best
fit'   the   same   cross   section   data.    As   shown   above   in
Fig.~\ref{Pb-skm}, $g$-folding model  evaluations gave quite good fits
to both the proton and neutron scattering cross sections.  Even better
fits to  the data were  obtained using phenomenological  optical model
potentials  (see   Fig~\ref{Pb-comp}).   Essentially  then   both  the
microscopic and phenomenological potentials yield relative motion wave
functions that  are the same asymptotically.   However, their relative
motion wave functions differ through the volume of the nucleus.

Whatever be the  (non-relativistic, spherical) nucleon-nucleus optical
model potential,    with the beam  direction as the $z$ axis, relative
motion wave functions  for   neutrons are written in partial wave form
\begin{equation}
\Psi_{\nu}({\bf k}, {\bf r}) 
= \frac{1}{k} \sum_{l, j} \sqrt{4\pi (2l + 1)}\; i^l f_{l, j}(kr)
\left\langle l \; \frac{1}{2} \; 0 \;\nu \left.\right| j \; \nu 
\right\rangle \mathcal{Y}_{l \frac{1}{2} j}^\nu (\Omega_r) ,
\end{equation}
where $\mathcal{Y}$  is the standard  tensor spherical harmonic  for a
spin-$\frac{1}{2}$ function.

\begin{figure}
\scalebox{0.6}{\includegraphics*{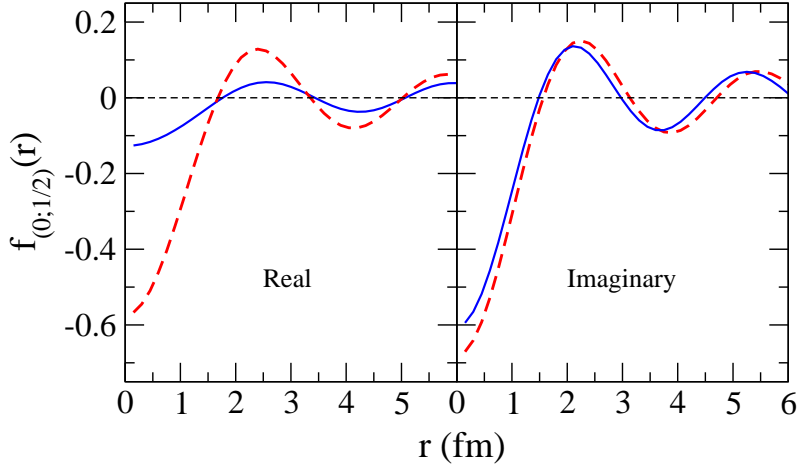}}
\caption{\label{s-waves}(Color  online) $s$-wave relative  motion wave
  functions for 40 MeV neutrons on $^{208}$Pb.  The real and imaginary
  parts are  as indicated with the  solid and dashed  lines giving the
  results  found  using  the  $g$-folding and  local  phenomenological
  optical model potentials respectively.}
\end{figure}

\begin{figure}
\scalebox{0.55}{\includegraphics*{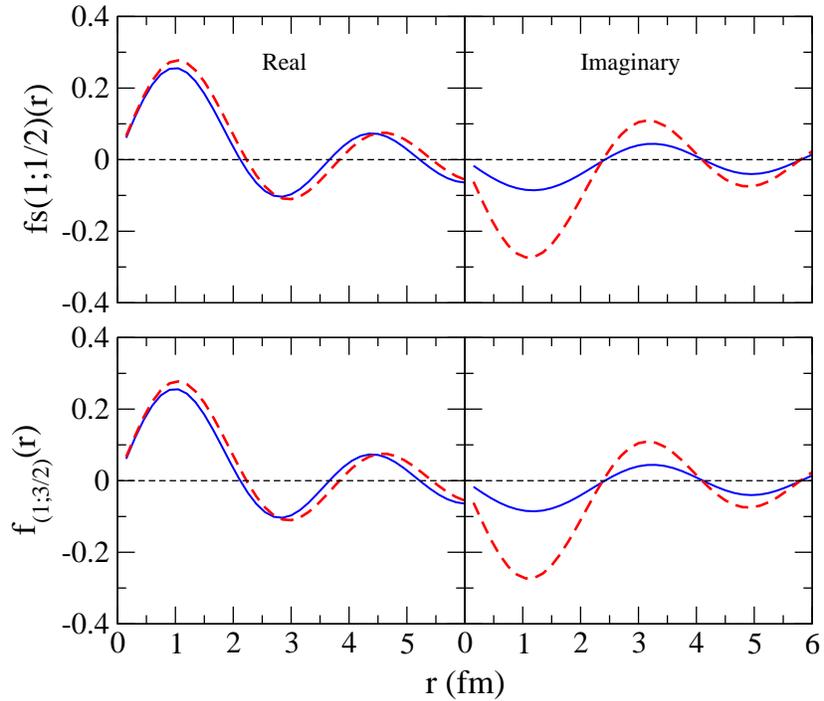}}
\caption{\label{p-waves}(Color  online) $p$-wave relative  motion wave
  functions for 40  MeV neutrons on $^{208}$Pb. The  results for spins
  $\frac{1}{2}$ and  $\frac{3}{2}$ are in the top  and bottom segments
  respectively. Other specifics are as for Fig.~\ref{s-waves}.}
\end{figure}

\begin{figure}
\scalebox{0.55}{\includegraphics*{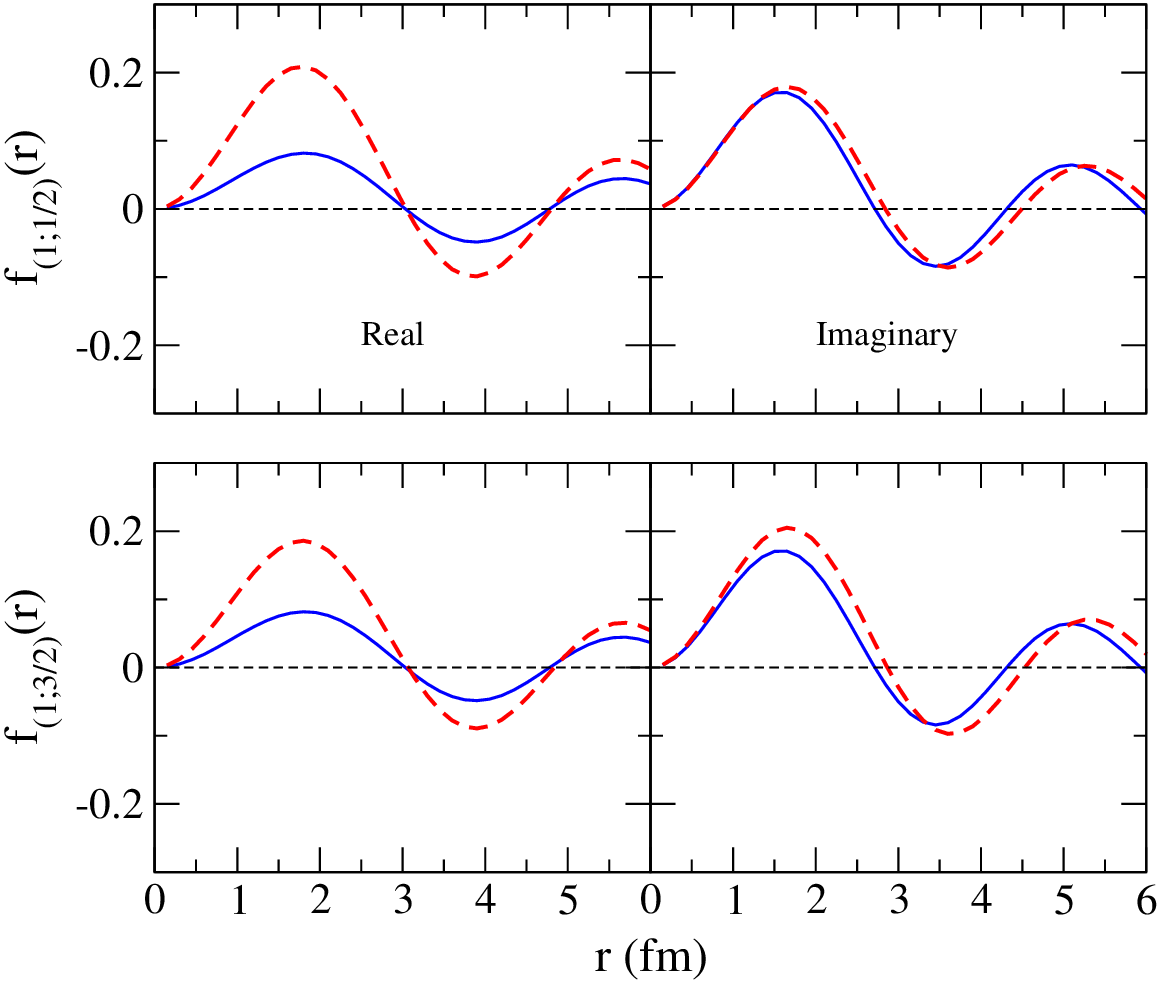}}
\caption{\label{d-waves}(Color  online) $d$-wave relative  motion wave
  functions for 40 MeV neutrons  on $^{208}$Pb.  The results for spins
  $\frac{3}{2}$ and  $\frac{5}{2}$ are in the top  and bottom segments
  respectively. Other specifics are as for Fig.~\ref{s-waves}.}
\end{figure}

Consider  the  case of  40  MeV  neutrons  elastically scattered  from
$^{208}$Pb.  The relative motion  wave functions for the $s$--, $p$--,
and  $d$--   partial  waves  are   displayed  in  Figs.~\ref{s-waves},
~\ref{p-waves},   and  \ref{d-waves}   respectively.   The   real  and
imaginary  parts of those  wave functions  are shown  on the  left and
right hand  sides of the figures.   The solid curves  portray the wave
functions generated  by the nonlocal,  microscopic $g$-folding optical
potential  while  those  found  using the  phenomenological  potential
\cite{Fi84} are  shown by the  dashed curves. All wave  functions were
obtained from  specific runs of  Raynal's updated version,  DWBA98, of
his code DWBA91~\cite{Ra91}.

The effect of nonlocality upon $s$-wave relative motion wave functions
through  the  nuclear  volume,  shown  in  Fig.~\ref{s-waves},  is  to
markedly decrease the real part while having a smaller decrease on the
imaginary part of  the wave function compared to  that found using the
(essentially phase  equivalent) local potential model.  The scaling is
not  the  overall 10-15\%  that  has  often  been assumed  (the  Perey
effect~\cite{Au65,Fi66}).

The $p$- and $d$- wave functions are similarly effected, though in the
case of the  $p$-waves, it is with the imaginary part  as given by the
code, that the large size reduction due to nonlocality occurs.

\begin{figure}[h]
\scalebox{0.65}{\includegraphics*{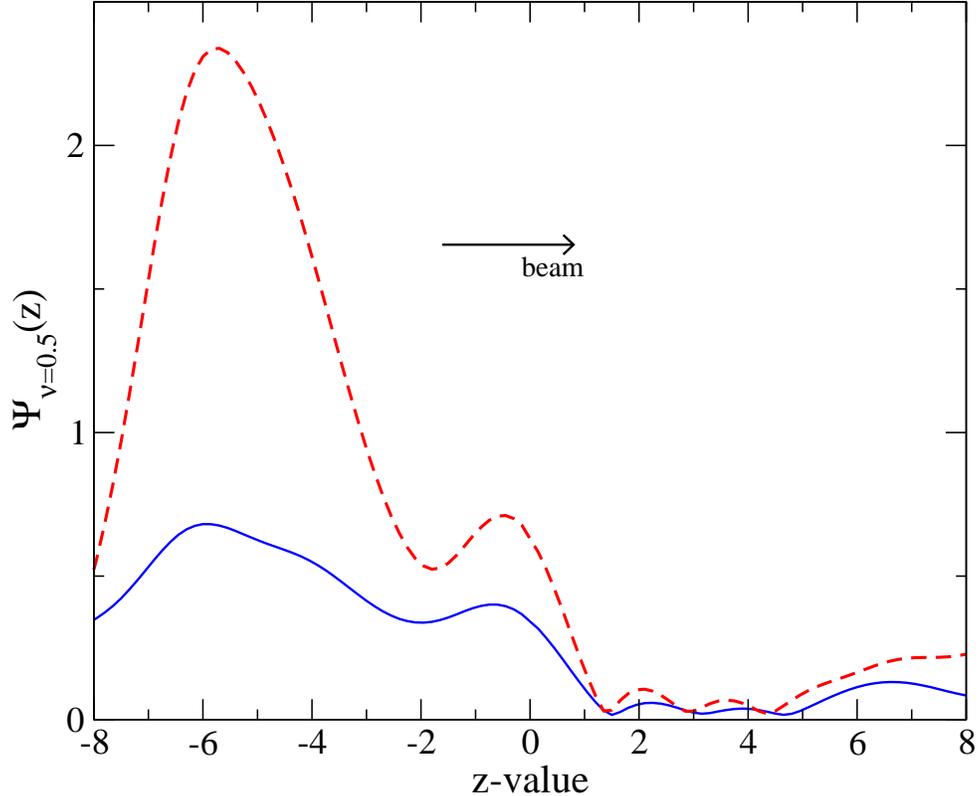}}
\caption{\label{Full-wv}(Color online) The  magnitudes of the complete
  scattering  waves  for  40  MeV  neutrons on  $^{208}$Pb  along  the
  $z$ axis.  The $g$-folding interaction  gave the result shown by the
  solid curve while that found from the phenomenological optical model
  is depicted by the dashed curve.}
\end{figure}

To emphasise the difference between the wave functions associated with
the  two  `phase  equivalent'  interactions  for 40  MeV  neutrons  on
$^{208}$Pb, the  partial wave sum  completed along the beam  line (for
which  $\Omega_r  = \pi$  for  $z  < 0$  and  0  thereafter, gave  the
magnitude of  the spin up  ($\nu = \frac{1}{2}$) scattering  waves as
displayed in Fig.~\ref{Full-wv}.

As  nuclear structure  models, by  virtue of  the fact  that  they are
models, do not precisely describe nuclei, there are mismatches between
observed  quantities  and  the  values  predicted.   Whether  that  be
$\gamma$-decay rates  or inelastic scattering cross  sections, a match
is  sought by using  effective transition  operators with  the ensuing
scaling of  calculated results  defined as `core  polarisation', which
are  considered in  the  next section.  Nevertheless, the  differences
obtained in the relative motion  wave functions when one considers the
effects in  scattering (Pauli blocking,  nonlocality, etc.) illustrate
the problem in describing nucleon-induced reactions and the correction
factors that may be  necessary \textit{a posteriori}. Such corrections
were postulated  as necessary in  a recent article  on coupled-channel
descriptions  of  reaction  cross  sections of  excited  states  which
ignores the effects  of Pauli and nonlocality in  the specification of
the effective interaction \cite{Ka09}.
 
\subsection{Core polarisation}

Core polarisation  enhancement of  calculated cross sections,  or form
factors, to replicate data is a measure of inadequacy of the structure
model   used.    Such    enhancement   should   be   momentum-transfer
dependent. That  makes using any specified core  polarisation from one
data set  not necessarily the correct  one for use  with another.  For
example,  to use  electromagnetic transition  rates  (essentially zero
momentum  quantities) to  specify a  core polarisation  for scattering
cross sections  typically specified for  a range of  momentum transfer
($\sim 0.2$ to $\sim 2.5$~fm$^{-1}$) is erroneous, as is the reverse.

\subsubsection{Core polarization from electromagnetic properties}

Consider the transition between the ground and first excited states in
$^6$Li for which the $B(E2)$ and longitudinal electron scattering form
factor are both  well known.  However, first we  need specify in brief
the shell models  to be used in evaluations  of both quantities.  More
details are given in Ref.~\cite{Ka97}.

\begin{figure}
\scalebox{0.8}{\includegraphics*{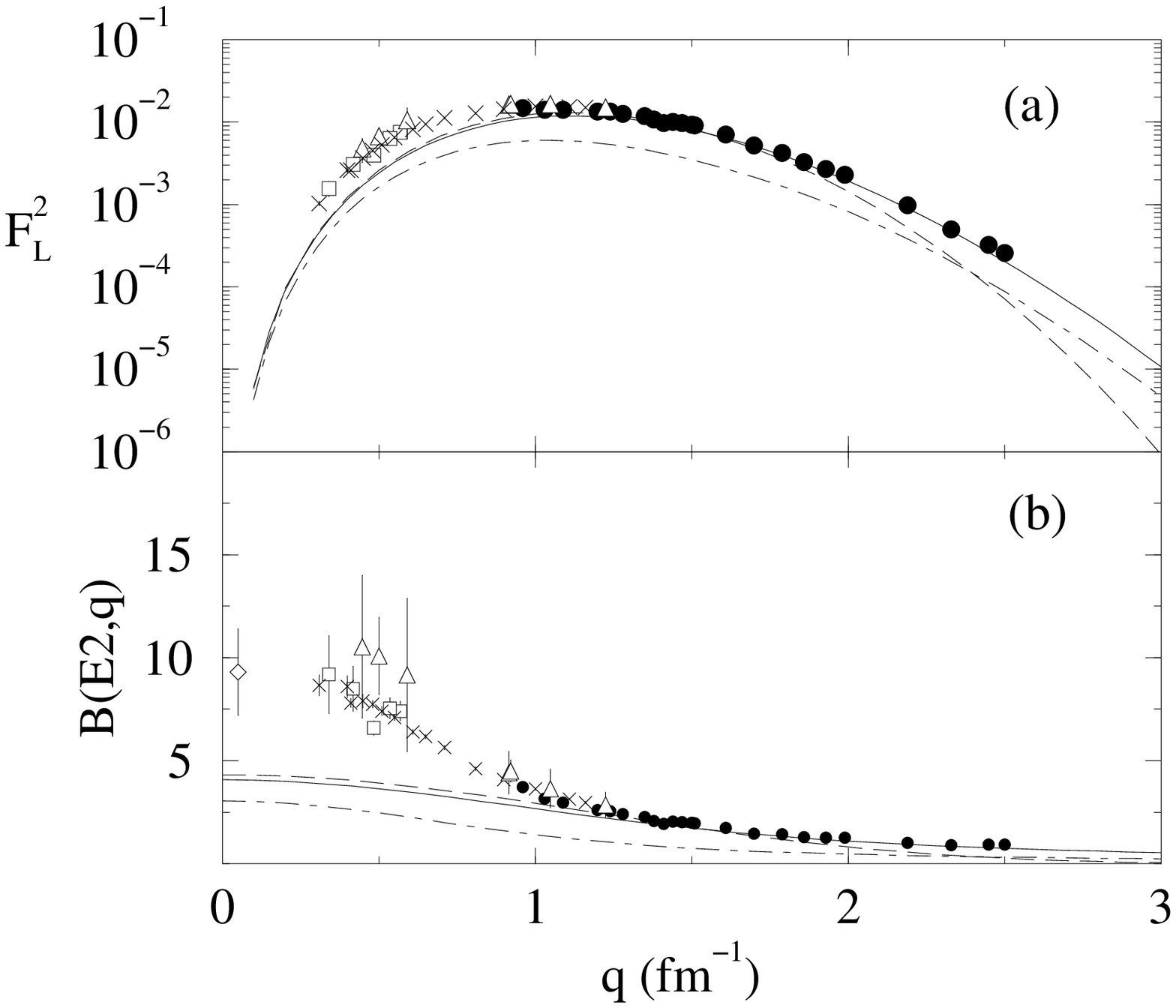}}
\caption{\label{elec-ff}    (a)   Longitudinal    inelastic   electron
  scattering form  factor to the $3^+;0$ (2.186~MeV)  state in $^6$Li,
  and (b) the $B(E2\downarrow,q)$  values, in units of $e^2$fm$^4$, as
  obtained from  the form factor  \cite{Ka97}.  The data  of Bergstrom
  \textit{et  al.}~\cite{Be79}  (circles),   of  Yen  \textit{et  al.}
  \cite{Ye74}  (squares),   of  Bergstrom  and   Tomusiak  \cite{Be76}
  (crosses), and  of Hutcheon  and Caplan \cite{Hu69}  (triangles) are
  compared    with   the   results    of   the    calculations.    The
  $B(E2\downarrow)$ value  from the associated  $\gamma$-decay rate is
  displayed by the diamond data point in (b).}
\end{figure}

Consider  shell  model  wave  functions within  the  $0\hbar  \omega$,
$(0+2)\hbar\omega$, and $(0+2+4)\hbar\omega$ model spaces.  The choice
of model space dictates the choice of interaction.  The ones used were
the  Cohen and  Kurath (6-16)2BME  interaction (CK)  for  the complete
$0\hbar\omega$  model space,  the  MK3W interaction  for the  complete
$(0+2)\hbar\omega$ shell  model space, and  the $G$-matrix interaction
of    Zheng    \textit{et    al.}~\cite{Zh95}   for    the    complete
$(0+2+4)\hbar\omega$  model  space.  The  shell  model wave  functions
found  with each  set were  used  to find  the longitudinal  inelastic
electron  scattering form  factor to  the $3^+;0$  (2.185  MeV) state,
amongst others~\cite{Ka97}. In Fig.~\ref{elec-ff}, the results found
using  the $(0+2+4)\hbar\omega$  wave  functions are  depicted by  the
solid  curves, those with  the $(0+2)\hbar\omega$  ones by  the dashed
curves,  and those  with the  $0\hbar\omega$ model  by  the dot-dashed
curves.  In the panel (a),  the data are those of Bergstrom \textit{et
  al.}~\cite{Be76}  (circles),  of  Yen \textit{et  al.}   \cite{Ye74}
(squares),  of  Bergstrom   and  Tomusiak~\cite{Be79}  (crosses),  and
Hutcheon  and  of Caplan~\cite{Hu69}  (triangles).   The form  factor,
calculated with all shell models,  is dominated by the $C2$ component,
while the $C4$ component is found  to be negligible. With the MK3W and
Zheng models,  the calculated results  reproduce the magnitude  of the
measured  form  factor above  1~fm$^{-1}$.   Both  have strength  from
transitions outside of the $0p$ shell which enhance the $C2$ strength.
Such  are  missing in  the  $0\hbar\omega$  model.  The  $(0+2+4)\hbar
\omega$  model structure  is most  favoured as  there is  almost exact
agreement with the data in  that region of momentum transfer. However,
the $B(E2)$  value for the  $\gamma$-decay of this $3^+_1;0$  state is
$9.3  \pm 2.1$~$e^2$fm$^4$,  and  the values  obtained by  calculation
using    the   three   models    of   structure    are   significantly
smaller~\cite{Ka97}.  So  far as the $\gamma$-decay  is concerned, all
calculations  require a substantial  renormalization to  reproduce the
measured  value.  That  is confirmed  by predictions  of  the electron
scattering form  factor at  low momentum transfer.   Below 1~fm$^{-1}$
all of  the calculated  results are less  than observation.   Yet that
degree  of renormalization  is not  suggested  by the  results of  the
calculations of  the form factor  at higher momentum  transfer.  While
this suggests that the internal  (nucleon) dynamics of the nucleus are
well described by the inclusion of higher $\hbar\omega$ excitations in
the  model space,  such  cannot  account for  the  asymptotics of  the
structure.   At large radii,  which most  influence scattering  at low
momentum  transfer,  the clustering  of  the  wave  functions are  not
reproduced  by   the  shell  model  in  which   up  to  $4\hbar\omega$
excitations  are included.   That is  also  why the  shell model  wave
functions also do  not give the correct quadrupole  moment for $^7$Li.
This deviation  of all  the calculated results  away from the  data is
illustrated further  in the (b)  panel of Fig.~\ref{elec-ff}  in which
the  $B(E2\downarrow,q)$ value for  the $3^+;0$  (2.186 MeV)  state in
$^6$Li is shown as a function of momentum transfer.  These values have
been determined from the measured and predicted longitudinal inelastic
scattering  form factors, achieved  by removing  from the  form factor
most  of the  dependence on  the  momentum transfer  according to  the
transformation  of  Brown,  Radhi,  and Wildenthal  \cite{Br83}.   The
$B(E2\downarrow)$  value is related  to the  associated $\gamma$-decay
which is given by the $q \sim 0$ intercept.

\subsubsection{Core polarization from radial transition form factors}

Core polarization  has been required  in DWIA (distorted  wave impulse
approximation)  analyses   of  pion  inelastic   scattering  to  match
calculated  cross  sections  to observed  data~\cite{Am83}.   Therein,
large  basis microscopic  models  of nuclear  structure  were used  to
specify the `collective' form  factors for inelastic scattering to the
$2_1^+$   states  in  $^{12}$C   and  $^{28}$Si.    The  $\pi$-nucleon
$t$-matrices input  were fixed by  fits to $\pi-N$  elastic scattering
phase shifts. All details are given in Ref.~\cite{Am83}.

The large basis models  of nuclear structure used in Ref.~\cite{Am83},
though  superseded now, suffice  for the  point to  be made.   For the
transition  in  $^{12}$C,   a  large  basis  projected  Hartree-Fock
calculation was  made.  For the  transition in $^{28}$Si,  large basis
Nilsson model wave functions were  used.  All single nucleon orbits to
$4\hbar\omega$  excitation  were  considered  and,  from  those  model
structures, transition one-body density matrices (OBDME) were defined.
Of those the dominant set of values were in the $p$-shell for $^{12}$C
and in the $sd$-shell for  $^{28}$Si. Those values are very similar to
what was found using the  conventional $0\hbar \omega$ shell model for
each nucleus.  However, the other  entries were not trivial and can be
considered  as  the  `core  polarization'  corrections  to  the  basic
$0\hbar\omega$ model values.

\begin{figure}
\scalebox{0.7}{\includegraphics*{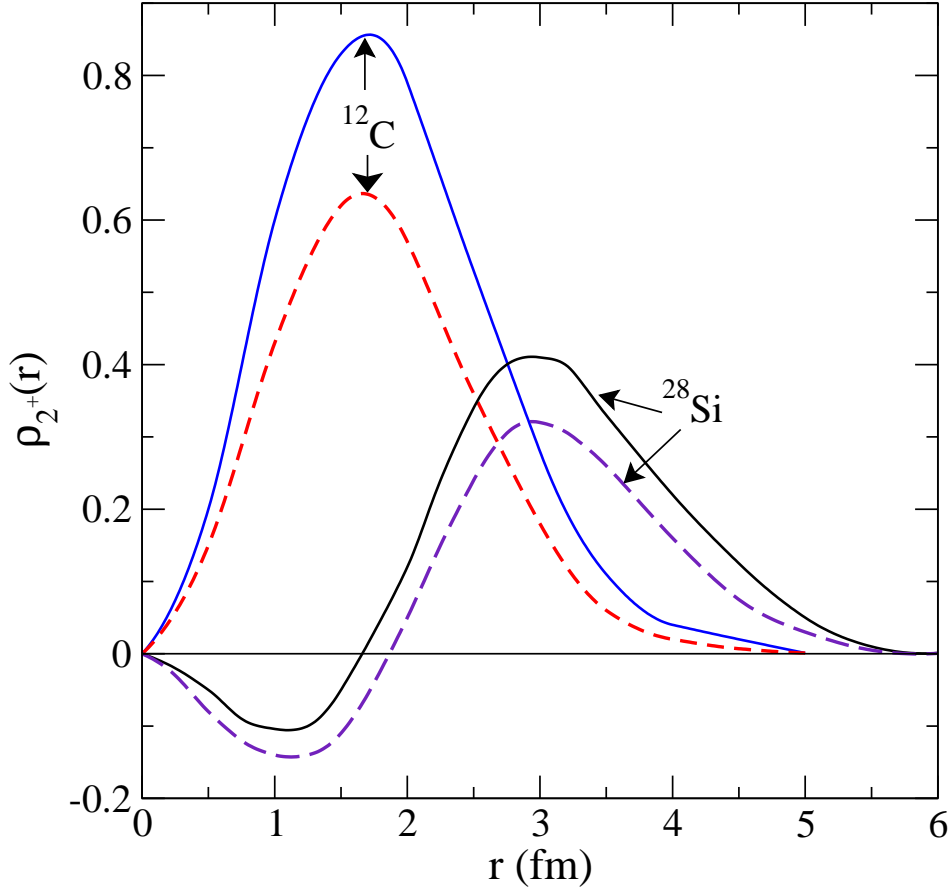}}
\caption{\label{pion-inel}(Color  online) Transition  radial densities
  for  the excitation  of the  $2_1^+$  (4.44 MeV)  state in  $^{12}$C
  (left) and of the $2_1^+$ (MeV) state in $^{28}$Si.}
\end{figure}

With  those OBDME, and  with harmonic  oscillator single  nucleon wave
functions,  radial  transition densities  for  the  excitation of  the
$2_1^+$ states from  the ground states were determined.   They are the
transition form factors  required in the DWIA evaluations  of the pion
(inelastic)  scattering  cross   sections.   These  radial  transition
densities for both nuclei are shown in Fig.~\ref{pion-inel}.

The solid curves are the transition densities found using the complete
set  of OBDME from  the large  basis wave  model wave  functions.  The
dashed  curves are  the densities  formed by  restricting the  sums of
OBDME to just the $p$-shell terms for $^{12}$C and just the $sd$-shell
entries for  $^{28}$Si.  Comparison of the two  densities for $^{12}$C
indicates that the `core polarization' enhances the density profile so
that  any cross  section  (essentially the  Fourier  transform of  the
transition density) would be  simply increased in magnitude consistent
with  the conventional  usage of  polarization charges.   In  fact the
increase is  a scale of  $\sim 1.4$ so  that a cross section  would be
increased by  a factor $\sim 2$.   However, that is not  the case with
$^{28}$Si.   Clearly   the  addition   of   the  `core   polarization'
contributions changes the profile  of the transition density, reducing
its  strength in  the region  $0\le  r \le  2$ fm  but increasing  the
strength at larger radii. The  Fourier transforms of the two then will
differ, and one can expect so also will the deduced cross sections.
 
\subsubsection{The $M_n/M_p$ approach}

There  are a  number of  studies, that  of Chien  and Khoa~\cite{Ch09}
being  a most recent  example, in  which neutron  scattering amplitude
scalings   were  linked   to  proton   ones   through  electromagnetic
properties.  Those  studies have used  the $M_n/M_p$ ratio  defined by
Bernstein,  Brown,  and  Madsen~\cite{Be81}  who recognised  that  the
combined  measurements  of  inelastic  hadron and  of  electromagnetic
excitation of  the same  states in  a nucleus could  provide a  way to
probe,  simultaneously,  the proton  and  neutron  excitations in  the
transition.  They formulated  a phenomenological  link to  specify the
ratio of multipole transition moments for protons ($M_p$) and neutrons
($M_n$), namely
\begin{equation}
\frac{M_n}{M_p} = \frac{b_p}{b_n}
\left[ \frac{\delta_h}{\delta_{em}}
\left( 1 + \frac{b_n}{b_p} \frac{N}{Z} \right)
- 1 \right] ,
\label{ratio}
\end{equation}
where for the  nucleus ($Z,N$), $\delta_h$ is a  deformation length to
be   determined  from  $(p,p')$   scattering,  $\delta_{em}$   is  the
electromagnetically  defined deformation  length, and  $b_n,  b_p$ are
interaction strengths of the  incident proton with target neutrons and
protons respectively.

The multipole transition elements themselves are defined by
\begin{equation}
M_{p(n)} = 
\left\langle J_f \left\| \sum_{i \in p(n)} r_i^\lambda Y_\lambda(\Omega_i) 
\right\| J_i \right\rangle
= \int_0^\infty \rho^{p(n)}_\lambda(r)\ r^{\lambda + 2}\; dr,
\label{Mpn}
\end{equation}
where $\rho^{p(n)}_\lambda(r)$  are transition densities  as described
in the preceding subsection and  $\lambda$ is the multipolarity of the
transition.

Usually the  $B(E2)$ values are used~\cite{Ma99,Iw08},  though in Ref.
\cite{Ch09}, it was  the $B(E2)$ values from the  $2_1^+$ state decays
from proton-rich $^{18}$Ne and  $^{20}$Mg to compare with their mirror
systems  in  the   neutron-rich  Oxygen  isotopes.   Iwasa  \textit{et
  al.}~\cite{Iw08} noted that there could be systematic uncertainties,
possibly caused  by insufficient knowledge of the  mechanism of proton
inelastic scattering.   As the  proton multipole transition  moment is
determined solely  by electromagnetic  data, the approach  negates any
contributions  to  $M_p$  from  the effective  $NN$  interaction.  The
process  is  not   a  reliable  way  to  assess   details  of  nuclear
structure. Likewise  it should not  be used to specify  $B(E2)$ values
from inelastic scattering cross sections from hydrogen targets, as has
been considered for the spectra of radioactive ions.

Though many of the problems with this approach have already discussed,
it is  worthwhile listing  them for emphasis.   First the  $B(E2)$ are
essentially  a zero  momentum transfer  property which,  if  a perfect
model of structure were available,  would depend upon, only the proton
transition density, with
\begin{equation}
B(E2) = \frac{1}{(2 J_i + 1)} M_p^2
\end{equation}
in units of  $e^2$fm$^4$. It is questionable therefore  to use this to
compare  data from  a  mirror  system, for  example  the $B(E2)$  from
$^{20}$Mg to that in $^{20}$O.  Of course, there are mirror symmetries
in  spectra  of   mirror  nuclei,  but  save  for   the  weak  Coulomb
displacement energies,  the symmetry is due to  charge independence of
the nuclear force.  However, at first order, the neutrons in a nucleus
do not contribute to $B(E2)$ values.   They do contribute in so far as
they  influence  proton distributions  through  the $NN$  interactions
being  stronger for  a  $pn$ pair  than  with like  nucleons, but  not
directly otherwise.

Hadron inelastic  scattering, of protons in  particular, have measured
cross sections usually  in a finite range of  momentum transfer values
that are  not very close  to zero.  That  is particularly so  with the
inverse kinematic  problem of radioactive ion  beams (RIBs) scattering
from hydrogen targets.   As noted previously, there is  a better match
in momentum transfer values if one compares electron form factors with
$(p,p')$  cross   sections,  but  experimental   data  from  electrons
scattering with  RIBs, either a cross  beam by or using  a SCRIT (self
contained radioactive ion target) experiment \cite{Su05} are yet to be
realised.

As stated earlier, perhaps the biggest problem of using  the $M_n/M_p$
ratio method is  that it assumes that the proton scattering transition
moment may be  solely determined by fits to  electromagnetic data. The
allowance for  specific nuclear interaction  effects, the role  of the
Pauli principle  in making  the actual transition  amplitudes creating
exchange  amplitudes which  often have  serious interference  with the
direct ones,  and the  effects of nuclear  distortion in  the relative
motion wave functions, are all missing in the identifications given in
Eqs.~(\ref{ratio}) and (\ref{Mpn}).  These reasons bring into question
the conclusions  made in  the recent article~\cite{Ch09}  dealing with
neutron  transition strengths  of $2^+_1$  states in  the neutron-rich
Oxygen isotopes.   That is especially  so since the assessment  of the
structure of those isotopes specified by a large basis shell model and
their use in analyses of the  scattering cross  sections of those RIBs  
from  hydrogen targets existed in the literature~\cite{Ka07}.

\section{Conclusions}

Of course, if only a set  of elastic scattering phase shifts that lead
to   a  quality  fit   to  elastic   scattering  and   total  reaction
cross-section data  are required, then  any prescription that  does so
can be used.  But usually the  elastic scattering data and its fit are
but a  prelude to what is sought.   By itself, one might  seek to make
conclusions about matter distributions in a nucleus from an assessment
of elastic  scattering data.  To do  so, however, one need  use a more
detailed theory of that scattering than usually has been the case.  At
least  the Pauli  principle and  the nonlocality  of the  $NA$ optical
potential  that it  creates  must  be taken  into  account, and  those
nonlocalities should  not be  approximated with some  local equivalent
form.  There is no  need to do so now as the  suite of programs DWBA98
and above, facilitate such analyses.  The effect of medium corrections
in the projectile nucleon -- target nucleon effective interactions can
be handled in  those programs as well, and  a utilitarian prescription
for those effective  interactions has been published~\cite{Am00}.  The
computer programs  that generate  the effective interactions  by those
means are available upon request.

Often the  associated relative  motion (distorted) wave  functions are
required, such  as in DWA  analyses of inelastic, charge  and particle
exchange  reaction data.   Those analyses  seek  conclusions regarding
structures of  excited states and  of processes involved.   An example
was  the use to  constrain double  beta decay  rates~\cite{Am07}.  The
problems   regarding  approximations   leading   to  effective   local
interaction forms of the optical potential are accentuated in all such
analyses  with the  radial properties  of the  distorted waves  from a
nonlocal  interaction  being   quite  different  through  the  nuclear
transition interaction  region (interior  and surface of  the nucleus)
from those of  an ``equivalent'' local optical potential  with which a
good fit to elastic scattering data was found. 

While this study has centred upon the problems associated with an $NA$
optical potential and its use in inelastic scattering analyses, many of
the same issues attend analyses of nucleus-nucleus scattering data and
conclusions about structure drawn from them.


\section*{Acknowledgments}
SK acknowledges support from the National Research Foundation of South
Africa.

\bibliography{Critiques}

\begin{thebibliography}{56}
\expandafter\ifx\csname natexlab\endcsname\relax\def\natexlab#1{#1}\fi
\expandafter\ifx\csname bibnamefont\endcsname\relax
  \def\bibnamefont#1{#1}\fi
\expandafter\ifx\csname bibfnamefont\endcsname\relax
  \def\bibfnamefont#1{#1}\fi
\expandafter\ifx\csname citenamefont\endcsname\relax
  \def\citenamefont#1{#1}\fi
\expandafter\ifx\csname url\endcsname\relax
  \def\url#1{\texttt{#1}}\fi
\expandafter\ifx\csname urlprefix\endcsname\relax\def\urlprefix{URL }\fi
\providecommand{\bibinfo}[2]{#2}
\providecommand{\eprint}[2][]{\url{#2}}

\bibitem[{\citenamefont{Bethe}(1940)}]{Be40}
\bibinfo{author}{\bibfnamefont{H.}~\bibnamefont{Bethe}},
  \bibinfo{journal}{Phys. Rev.} \textbf{\bibinfo{volume}{57}},
  \bibinfo{pages}{1125} (\bibinfo{year}{1940}).

\bibitem[{\citenamefont{Kerman et~al.}(1959)\citenamefont{Kerman, McManus, and
  Thaler}}]{Ke59}
\bibinfo{author}{\bibfnamefont{A.~K.} \bibnamefont{Kerman}},
  \bibinfo{author}{\bibfnamefont{H.}~\bibnamefont{McManus}}, \bibnamefont{and}
  \bibinfo{author}{\bibfnamefont{R.~M.} \bibnamefont{Thaler}},
  \bibinfo{journal}{Ann. Phys. (N.Y.)} \textbf{\bibinfo{volume}{8}},
  \bibinfo{pages}{551} (\bibinfo{year}{1959}).

\bibitem[{\citenamefont{Amos et~al.}(2000)\citenamefont{Amos, Dortmans, von
  Geramb, Karataglidis, and Raynal}}]{Am00}
\bibinfo{author}{\bibfnamefont{K.}~\bibnamefont{Amos}},
  \bibinfo{author}{\bibfnamefont{P.~J.} \bibnamefont{Dortmans}},
  \bibinfo{author}{\bibfnamefont{H.~V.} \bibnamefont{von Geramb}},
  \bibinfo{author}{\bibfnamefont{S.}~\bibnamefont{Karataglidis}},
  \bibnamefont{and} \bibinfo{author}{\bibfnamefont{J.}~\bibnamefont{Raynal}},
  \bibinfo{journal}{Adv. in Nucl. Phys.} \textbf{\bibinfo{volume}{25}},
  \bibinfo{pages}{275} (\bibinfo{year}{2000}), \bibinfo{note}{(and references
  contained therein)}.

\bibitem[{\citenamefont{Jeukenne et~al.}(1976)\citenamefont{Jeukenne, Lejeunne,
  and Mahaux}}]{Je76}
\bibinfo{author}{\bibfnamefont{J.~P.} \bibnamefont{Jeukenne}},
  \bibinfo{author}{\bibfnamefont{A.}~\bibnamefont{Lejeunne}}, \bibnamefont{and}
  \bibinfo{author}{\bibfnamefont{C.}~\bibnamefont{Mahaux}},
  \bibinfo{journal}{Phys. Rep.} \textbf{\bibinfo{volume}{25}},
  \bibinfo{pages}{83} (\bibinfo{year}{1976}).

\bibitem[{\citenamefont{Becheva et~al.}(2006)}]{Be06}
\bibinfo{author}{\bibfnamefont{E.}~\bibnamefont{Becheva}} \bibnamefont{et~al.},
  \bibinfo{journal}{Phys. Rev. Lett.} \textbf{\bibinfo{volume}{96}},
  \bibinfo{pages}{012501} (\bibinfo{year}{2006}).

\bibitem[{\citenamefont{Lun et~al.}(1998)\citenamefont{Lun, Ebersp{\"a}cher,
  Amos, Scheid, and Buckman}}]{Lu98}
\bibinfo{author}{\bibfnamefont{D.~R.} \bibnamefont{Lun}},
  \bibinfo{author}{\bibfnamefont{M.}~\bibnamefont{Ebersp{\"a}cher}},
  \bibinfo{author}{\bibfnamefont{K.}~\bibnamefont{Amos}},
  \bibinfo{author}{\bibfnamefont{W.}~\bibnamefont{Scheid}}, \bibnamefont{and}
  \bibinfo{author}{\bibfnamefont{S.~J.} \bibnamefont{Buckman}},
  \bibinfo{journal}{Phys. Rev. A} \textbf{\bibinfo{volume}{58}},
  \bibinfo{pages}{4993} (\bibinfo{year}{1998}).

\bibitem[{\citenamefont{Lovell and Amos}(2000)}]{Lo00}
\bibinfo{author}{\bibfnamefont{A.}~\bibnamefont{Lovell}} \bibnamefont{and}
  \bibinfo{author}{\bibfnamefont{K.}~\bibnamefont{Amos}},
  \bibinfo{journal}{Phys. Rev. C} \textbf{\bibinfo{volume}{62}},
  \bibinfo{pages}{064614} (\bibinfo{year}{2000}).

\bibitem[{\citenamefont{Meyer et~al.}(1983)\citenamefont{Meyer, Schwandt,
  Jacobs, and Hall}}]{Me83}
\bibinfo{author}{\bibfnamefont{H.~O.} \bibnamefont{Meyer}},
  \bibinfo{author}{\bibfnamefont{P.}~\bibnamefont{Schwandt}},
  \bibinfo{author}{\bibfnamefont{W.~W.} \bibnamefont{Jacobs}},
  \bibnamefont{and} \bibinfo{author}{\bibfnamefont{J.~R.} \bibnamefont{Hall}},
  \bibinfo{journal}{Phys. Rev. C} \textbf{\bibinfo{volume}{27}},
  \bibinfo{pages}{459} (\bibinfo{year}{1983}).

\bibitem[{\citenamefont{Allen et~al.}(1994)\citenamefont{Allen, Amos, and
  Dortmans}}]{Al94}
\bibinfo{author}{\bibfnamefont{L.~J.} \bibnamefont{Allen}},
  \bibinfo{author}{\bibfnamefont{K.}~\bibnamefont{Amos}}, \bibnamefont{and}
  \bibinfo{author}{\bibfnamefont{P.~J.} \bibnamefont{Dortmans}},
  \bibinfo{journal}{Phys. Rev. C} \textbf{\bibinfo{volume}{49}},
  \bibinfo{pages}{2177} (\bibinfo{year}{1994}).

\bibitem[{\citenamefont{Stepansov et~al.}(2002)}]{St02}
\bibinfo{author}{\bibfnamefont{S.~V.} \bibnamefont{Stepansov}}
  \bibnamefont{et~al.}, \bibinfo{journal}{Phys. Letts.}
  \textbf{\bibinfo{volume}{B 542}}, \bibinfo{pages}{35} (\bibinfo{year}{2002}).

\bibitem[{\citenamefont{Deb and Amos}(2000)}]{De00}
\bibinfo{author}{\bibfnamefont{P.~K.} \bibnamefont{Deb}} \bibnamefont{and}
  \bibinfo{author}{\bibfnamefont{K.}~\bibnamefont{Amos}},
  \bibinfo{journal}{Phys. Rev. C} \textbf{\bibinfo{volume}{62}},
  \bibinfo{pages}{024605} (\bibinfo{year}{2000}).

\bibitem[{\citenamefont{Arndt et~al.}(1983)\citenamefont{Arndt, Roper, Bryan,
  Clark, ver West, and Signell}}]{Ar83}
\bibinfo{author}{\bibfnamefont{R.~A.} \bibnamefont{Arndt}},
  \bibinfo{author}{\bibfnamefont{L.~D.} \bibnamefont{Roper}},
  \bibinfo{author}{\bibfnamefont{R.~A.} \bibnamefont{Bryan}},
  \bibinfo{author}{\bibfnamefont{R.~B.} \bibnamefont{Clark}},
  \bibinfo{author}{\bibfnamefont{B.~J.} \bibnamefont{ver West}},
  \bibnamefont{and} \bibinfo{author}{\bibfnamefont{P.}~\bibnamefont{Signell}},
  \bibinfo{journal}{Phys. Rev. D} \textbf{\bibinfo{volume}{28}},
  \bibinfo{pages}{97} (\bibinfo{year}{1983}).

\bibitem[{\citenamefont{Nagels et~al.}(1978)\citenamefont{Nagels, Rijken, and
  de~Swart}}]{Na78}
\bibinfo{author}{\bibfnamefont{M.~M.} \bibnamefont{Nagels}},
  \bibinfo{author}{\bibfnamefont{T.~A.} \bibnamefont{Rijken}},
  \bibnamefont{and} \bibinfo{author}{\bibfnamefont{J.~J.}
  \bibnamefont{de~Swart}}, \bibinfo{journal}{Phys. Rev. D}
  \textbf{\bibinfo{volume}{17}}, \bibinfo{pages}{768} (\bibinfo{year}{1978}).

\bibitem[{\citenamefont{Lacombe et~al.}(1980)\citenamefont{Lacombe, Loiseau,
  Richard, Mau, C{\^o}t{\'e}, Pir{\'e}s, and de~Tourreil}}]{La80}
\bibinfo{author}{\bibfnamefont{M.}~\bibnamefont{Lacombe}},
  \bibinfo{author}{\bibfnamefont{B.}~\bibnamefont{Loiseau}},
  \bibinfo{author}{\bibfnamefont{J.~M.} \bibnamefont{Richard}},
  \bibinfo{author}{\bibfnamefont{R.~V.} \bibnamefont{Mau}},
  \bibinfo{author}{\bibfnamefont{J.}~\bibnamefont{C{\^o}t{\'e}}},
  \bibinfo{author}{\bibfnamefont{P.}~\bibnamefont{Pir{\'e}s}},
  \bibnamefont{and}
  \bibinfo{author}{\bibfnamefont{R.}~\bibnamefont{de~Tourreil}},
  \bibinfo{journal}{Phys. Rev. C} \textbf{\bibinfo{volume}{21}},
  \bibinfo{pages}{861} (\bibinfo{year}{1980}).

\bibitem[{\citenamefont{Machleidt et~al.}(1987)\citenamefont{Machleidt,
  Holinde, and Elster}}]{Ma87}
\bibinfo{author}{\bibfnamefont{R.}~\bibnamefont{Machleidt}},
  \bibinfo{author}{\bibfnamefont{K.}~\bibnamefont{Holinde}}, \bibnamefont{and}
  \bibinfo{author}{\bibfnamefont{C.}~\bibnamefont{Elster}},
  \bibinfo{journal}{Phys. Rep.} \textbf{\bibinfo{volume}{149}},
  \bibinfo{pages}{1} (\bibinfo{year}{1987}).

\bibitem[{\citenamefont{von Geramb}(1979)}]{Ge79}
\bibinfo{editor}{\bibfnamefont{H.~V.} \bibnamefont{von Geramb}}, ed.,
  \emph{\bibinfo{title}{Microscopic Optical Potentials, Proc. of the Hamburg
  Topical Workshop}}, vol.~\bibinfo{volume}{89} of
  \emph{\bibinfo{series}{Lecture Notes in Physics}}
  (\bibinfo{publisher}{Springer-Verlag}, \bibinfo{address}{Berlin},
  \bibinfo{year}{1979}).

\bibitem[{\citenamefont{Schwandt}(1983)}]{Sc83}
\bibinfo{author}{\bibfnamefont{P.}~\bibnamefont{Schwandt}}, in
  \emph{\bibinfo{booktitle}{Proc. Int. Workshop on Medium Energy Nucleons in
  Nuclei}}, edited by \bibinfo{editor}{\bibfnamefont{H.~O.}
  \bibnamefont{Meyer}} (\bibinfo{publisher}{AIP}, \bibinfo{address}{New York},
  \bibinfo{year}{1983}), vol.~\bibinfo{volume}{97} of
  \emph{\bibinfo{series}{AIP Conf. Proc.}}

\bibitem[{\citenamefont{Ray et~al.}(1992)\citenamefont{Ray, Hoffmann, and
  Coker}}]{Ra92}
\bibinfo{author}{\bibfnamefont{L.}~\bibnamefont{Ray}},
  \bibinfo{author}{\bibfnamefont{G.~W.} \bibnamefont{Hoffmann}},
  \bibnamefont{and} \bibinfo{author}{\bibfnamefont{W.~R.} \bibnamefont{Coker}},
  \bibinfo{journal}{Phys. Rep.} \textbf{\bibinfo{volume}{212}},
  \bibinfo{pages}{223} (\bibinfo{year}{1992}).

\bibitem[{\citenamefont{Satchler}(1983)}]{Sa83}
\bibinfo{author}{\bibfnamefont{G.~R.} \bibnamefont{Satchler}},
  \emph{\bibinfo{title}{Direct Nuclear Reactions}}, International series of
  monographs on physics (\bibinfo{publisher}{Clarendon}, \bibinfo{address}{New
  York}, \bibinfo{year}{1983}), \bibinfo{edition}{68th} ed.

\bibitem[{\citenamefont{Feshbach}(1992)}]{Fe92}
\bibinfo{author}{\bibfnamefont{H.}~\bibnamefont{Feshbach}},
  \emph{\bibinfo{title}{Theoretical Nuclear Physics: Nuclear Reactions}}
  (\bibinfo{publisher}{Wiley}, \bibinfo{address}{London},
  \bibinfo{year}{1992}).

\bibitem[{\citenamefont{Hodgson}(1994)}]{Ho94}
\bibinfo{author}{\bibfnamefont{P.~E.} \bibnamefont{Hodgson}},
  \emph{\bibinfo{title}{The Nucleon Optical Potential}}
  (\bibinfo{publisher}{World Scientific}, \bibinfo{address}{Singapore},
  \bibinfo{year}{1994}).

\bibitem[{\citenamefont{Ray}(1990)}]{Ra90}
\bibinfo{author}{\bibfnamefont{L.}~\bibnamefont{Ray}}, \bibinfo{journal}{Phys.
  Rev. C} \textbf{\bibinfo{volume}{41}}, \bibinfo{pages}{2816}
  (\bibinfo{year}{1990}).

\bibitem[{\citenamefont{Sakaguchi et~al.}(1982)}]{Sa82}
\bibinfo{author}{\bibfnamefont{H.}~\bibnamefont{Sakaguchi}}
  \bibnamefont{et~al.}, \bibinfo{journal}{Phys. Rev. C}
  \textbf{\bibinfo{volume}{26}}, \bibinfo{pages}{944} (\bibinfo{year}{1982}).

\bibitem[{\citenamefont{Karataglidis et~al.}(2002)\citenamefont{Karataglidis,
  Amos, Brown, and Deb}}]{Ka02}
\bibinfo{author}{\bibfnamefont{S.}~\bibnamefont{Karataglidis}},
  \bibinfo{author}{\bibfnamefont{K.}~\bibnamefont{Amos}},
  \bibinfo{author}{\bibfnamefont{B.~A.} \bibnamefont{Brown}}, \bibnamefont{and}
  \bibinfo{author}{\bibfnamefont{P.~K.} \bibnamefont{Deb}},
  \bibinfo{journal}{Phys. Rev. C} \textbf{\bibinfo{volume}{65}},
  \bibinfo{pages}{044306} (\bibinfo{year}{2002}).

\bibitem[{\citenamefont{Ibaraki et~al.}(2000)}]{Ib00}
\bibinfo{author}{\bibfnamefont{M.}~\bibnamefont{Ibaraki}} \bibnamefont{et~al.},
  \bibinfo{journal}{Nucl. Instrum. and Meth. in Phys. Res.}
  \textbf{\bibinfo{volume}{A 446}}, \bibinfo{pages}{536}
  (\bibinfo{year}{2000}).

\bibitem[{\citenamefont{van Oers et~al.}(1974)}]{vO74}
\bibinfo{author}{\bibfnamefont{W.~T.~H.} \bibnamefont{van Oers}}
  \bibnamefont{et~al.}, \bibinfo{journal}{Phys. Rev. C}
  \textbf{\bibinfo{volume}{10}}, \bibinfo{pages}{307} (\bibinfo{year}{1974}).

\bibitem[{\citenamefont{Finlay et~al.}(1984)\citenamefont{Finlay, Annand,
  Cheema, Rapaport, and Dietrich}}]{Fi84}
\bibinfo{author}{\bibfnamefont{R.~W.} \bibnamefont{Finlay}},
  \bibinfo{author}{\bibfnamefont{J.~R.~M.} \bibnamefont{Annand}},
  \bibinfo{author}{\bibfnamefont{T.~S.} \bibnamefont{Cheema}},
  \bibinfo{author}{\bibfnamefont{J.}~\bibnamefont{Rapaport}}, \bibnamefont{and}
  \bibinfo{author}{\bibfnamefont{F.~S.} \bibnamefont{Dietrich}},
  \bibinfo{journal}{Phys. Rev. C} \textbf{\bibinfo{volume}{30}},
  \bibinfo{pages}{796} (\bibinfo{year}{1984}).

\bibitem[{\citenamefont{Amos et~al.}(2003)\citenamefont{Amos, Canton, Pisent,
  Svenne, and van~der Knijff}}]{Am03}
\bibinfo{author}{\bibfnamefont{K.}~\bibnamefont{Amos}},
  \bibinfo{author}{\bibfnamefont{L.}~\bibnamefont{Canton}},
  \bibinfo{author}{\bibfnamefont{G.}~\bibnamefont{Pisent}},
  \bibinfo{author}{\bibfnamefont{J.~P.} \bibnamefont{Svenne}},
  \bibnamefont{and} \bibinfo{author}{\bibfnamefont{D.}~\bibnamefont{van~der
  Knijff}}, \bibinfo{journal}{Nucl.\ Phys.} \textbf{\bibinfo{volume}{A728}},
  \bibinfo{pages}{65} (\bibinfo{year}{2003}).

\bibitem[{\citenamefont{Canton et~al.}(2006)\citenamefont{Canton, Pisent,
  Svenne, Amos, and Karataglidis}}]{Ca06}
\bibinfo{author}{\bibfnamefont{L.}~\bibnamefont{Canton}},
  \bibinfo{author}{\bibfnamefont{G.}~\bibnamefont{Pisent}},
  \bibinfo{author}{\bibfnamefont{J.~P.} \bibnamefont{Svenne}},
  \bibinfo{author}{\bibfnamefont{K.}~\bibnamefont{Amos}}, \bibnamefont{and}
  \bibinfo{author}{\bibfnamefont{S.}~\bibnamefont{Karataglidis}},
  \bibinfo{journal}{Phys. Rev. Lett.} \textbf{\bibinfo{volume}{96}},
  \bibinfo{pages}{072502} (\bibinfo{year}{2006}).

\bibitem[{\citenamefont{Amos et~al.}(2005)\citenamefont{Amos, Karataglidis,
  van~der Knijff, Canton, Pisent, and Svenne}}]{Am05}
\bibinfo{author}{\bibfnamefont{K.}~\bibnamefont{Amos}},
  \bibinfo{author}{\bibfnamefont{S.}~\bibnamefont{Karataglidis}},
  \bibinfo{author}{\bibfnamefont{D.}~\bibnamefont{van~der Knijff}},
  \bibinfo{author}{\bibfnamefont{L.}~\bibnamefont{Canton}},
  \bibinfo{author}{\bibfnamefont{G.}~\bibnamefont{Pisent}}, \bibnamefont{and}
  \bibinfo{author}{\bibfnamefont{J.~P.} \bibnamefont{Svenne}},
  \bibinfo{journal}{Phys. Rev. C} \textbf{\bibinfo{volume}{72}},
  \bibinfo{pages}{064604} (\bibinfo{year}{2005}).

\bibitem[{\citenamefont{von Geramb et~al.}(1975)\citenamefont{von Geramb, Amos,
  Sprickmann, Kn{\"o}pfle, Rogge, Ingham, and Mayer-B{\"o}ricke}}]{Ge75}
\bibinfo{author}{\bibfnamefont{H.~V.} \bibnamefont{von Geramb}},
  \bibinfo{author}{\bibfnamefont{K.}~\bibnamefont{Amos}},
  \bibinfo{author}{\bibfnamefont{R.}~\bibnamefont{Sprickmann}},
  \bibinfo{author}{\bibfnamefont{K.~T.} \bibnamefont{Kn{\"o}pfle}},
  \bibinfo{author}{\bibfnamefont{M.}~\bibnamefont{Rogge}},
  \bibinfo{author}{\bibfnamefont{D.}~\bibnamefont{Ingham}}, \bibnamefont{and}
  \bibinfo{author}{\bibfnamefont{C.}~\bibnamefont{Mayer-B{\"o}ricke}},
  \bibinfo{journal}{Phys. Rev. C} \textbf{\bibinfo{volume}{12}},
  \bibinfo{pages}{1697} (\bibinfo{year}{1975}).

\bibitem[{\citenamefont{Tostevin et~al.}(2001)\citenamefont{Tostevin, Nunes,
  and Thompson}}]{To01}
\bibinfo{author}{\bibfnamefont{J.~A.} \bibnamefont{Tostevin}},
  \bibinfo{author}{\bibfnamefont{F.~M.} \bibnamefont{Nunes}}, \bibnamefont{and}
  \bibinfo{author}{\bibfnamefont{I.~J.} \bibnamefont{Thompson}},
  \bibinfo{journal}{Phys. Rev. C} \textbf{\bibinfo{volume}{63}},
  \bibinfo{pages}{024617} (\bibinfo{year}{2001}).

\bibitem[{\citenamefont{Deltuva et~al.}(2007)\citenamefont{Deltuva, Moro,
  Cravo, Nunes, and Fonseca}}]{De07}
\bibinfo{author}{\bibfnamefont{A.}~\bibnamefont{Deltuva}},
  \bibinfo{author}{\bibfnamefont{A.~M.} \bibnamefont{Moro}},
  \bibinfo{author}{\bibfnamefont{E.}~\bibnamefont{Cravo}},
  \bibinfo{author}{\bibfnamefont{F.~M.} \bibnamefont{Nunes}}, \bibnamefont{and}
  \bibinfo{author}{\bibfnamefont{A.~C.} \bibnamefont{Fonseca}},
  \bibinfo{journal}{Phys. Rev. C} \textbf{\bibinfo{volume}{76}},
  \bibinfo{pages}{064602} (\bibinfo{year}{2007}).

\bibitem[{\citenamefont{Skaza et~al.}(2005)}]{Sk05}
\bibinfo{author}{\bibfnamefont{F.}~\bibnamefont{Skaza}} \bibnamefont{et~al.},
  \bibinfo{journal}{Phys. Lett.} \textbf{\bibinfo{volume}{619B}},
  \bibinfo{pages}{92} (\bibinfo{year}{2005}).

\bibitem[{\citenamefont{Mackintosh and Keeley}(2010)}]{Ma10}
\bibinfo{author}{\bibfnamefont{R.~S.} \bibnamefont{Mackintosh}}
  \bibnamefont{and} \bibinfo{author}{\bibfnamefont{N.}~\bibnamefont{Keeley}},
  \bibinfo{journal}{Phys. Rev. C} \textbf{\bibinfo{volume}{81}},
  \bibinfo{pages}{034612} (\bibinfo{year}{2010}).

\bibitem[{\citenamefont{Khan et~al.}(2001)}]{Kh01}
\bibinfo{author}{\bibfnamefont{E.}~\bibnamefont{Khan}} \bibnamefont{et~al.},
  \bibinfo{journal}{Nucl. Phys.} \textbf{\bibinfo{volume}{A694}},
  \bibinfo{pages}{103} (\bibinfo{year}{2001}).

\bibitem[{\citenamefont{Deb et~al.}(2005)\citenamefont{Deb, Clark, Hama, Amos,
  Karataglidis, and Cooper}}]{De05}
\bibinfo{author}{\bibfnamefont{P.~K.} \bibnamefont{Deb}},
  \bibinfo{author}{\bibfnamefont{B.~C.} \bibnamefont{Clark}},
  \bibinfo{author}{\bibfnamefont{S.}~\bibnamefont{Hama}},
  \bibinfo{author}{\bibfnamefont{K.}~\bibnamefont{Amos}},
  \bibinfo{author}{\bibfnamefont{S.}~\bibnamefont{Karataglidis}},
  \bibnamefont{and} \bibinfo{author}{\bibfnamefont{E.~D.}
  \bibnamefont{Cooper}}, \bibinfo{journal}{Phys. Rev. C}
  \textbf{\bibinfo{volume}{72}}, \bibinfo{pages}{014608}
  (\bibinfo{year}{2005}).

\bibitem[{\citenamefont{Raynal}(1981)}]{Ra91}
\bibinfo{author}{\bibfnamefont{J.}~\bibnamefont{Raynal}},
  \emph{\bibinfo{title}{computer code dwba70}} (\bibinfo{year}{1981}),
  \bibinfo{note}{(NEA 1209/02)}.

\bibitem[{\citenamefont{Austern}(1965)}]{Au65}
\bibinfo{author}{\bibfnamefont{N.}~\bibnamefont{Austern}},
  \bibinfo{journal}{Phys. Rev.} \textbf{\bibinfo{volume}{137}},
  \bibinfo{pages}{B752} (\bibinfo{year}{1965}).

\bibitem[{\citenamefont{Fiedeldey}(1966)}]{Fi66}
\bibinfo{author}{\bibfnamefont{H.}~\bibnamefont{Fiedeldey}},
  \bibinfo{journal}{Nucl. Phys.} \textbf{\bibinfo{volume}{77}},
  \bibinfo{pages}{149} (\bibinfo{year}{1966}).

\bibitem[{\citenamefont{Kawano et~al.}(2009)\citenamefont{Kawano, Talou, Lynn,
  Chadwick, and Madland}}]{Ka09}
\bibinfo{author}{\bibfnamefont{T.}~\bibnamefont{Kawano}},
  \bibinfo{author}{\bibfnamefont{P.}~\bibnamefont{Talou}},
  \bibinfo{author}{\bibfnamefont{J.~E.} \bibnamefont{Lynn}},
  \bibinfo{author}{\bibfnamefont{M.~B.} \bibnamefont{Chadwick}},
  \bibnamefont{and} \bibinfo{author}{\bibfnamefont{D.~G.}
  \bibnamefont{Madland}}, \bibinfo{journal}{Phys. Rev. C}
  \textbf{\bibinfo{volume}{80}}, \bibinfo{pages}{024611}
  (\bibinfo{year}{2009}).

\bibitem[{\citenamefont{Karataglidis et~al.}(1997)\citenamefont{Karataglidis,
  Brown, Amos, and Dortmans}}]{Ka97}
\bibinfo{author}{\bibfnamefont{S.}~\bibnamefont{Karataglidis}},
  \bibinfo{author}{\bibfnamefont{B.~A.} \bibnamefont{Brown}},
  \bibinfo{author}{\bibfnamefont{K.}~\bibnamefont{Amos}}, \bibnamefont{and}
  \bibinfo{author}{\bibfnamefont{P.~J.} \bibnamefont{Dortmans}},
  \bibinfo{journal}{Phys. Rev. C} \textbf{\bibinfo{volume}{55}},
  \bibinfo{pages}{2826} (\bibinfo{year}{1997}).

\bibitem[{\citenamefont{Zheng et~al.}(1995)\citenamefont{Zheng, Barrett, Vary,
  Haxton, and Song}}]{Zh95}
\bibinfo{author}{\bibfnamefont{D.~C.} \bibnamefont{Zheng}},
  \bibinfo{author}{\bibfnamefont{B.~R.} \bibnamefont{Barrett}},
  \bibinfo{author}{\bibfnamefont{J.~P.} \bibnamefont{Vary}},
  \bibinfo{author}{\bibfnamefont{W.~C.} \bibnamefont{Haxton}},
  \bibnamefont{and} \bibinfo{author}{\bibfnamefont{C.-L.} \bibnamefont{Song}},
  \bibinfo{journal}{Phys. Rev. C} \textbf{\bibinfo{volume}{52}},
  \bibinfo{pages}{2488} (\bibinfo{year}{1995}).

\bibitem[{\citenamefont{Bergstrom et~al.}(1979)\citenamefont{Bergstrom,
  Deutschmann, and Neuhausen}}]{Be79}
\bibinfo{author}{\bibfnamefont{J.~C.} \bibnamefont{Bergstrom}},
  \bibinfo{author}{\bibfnamefont{U.}~\bibnamefont{Deutschmann}},
  \bibnamefont{and}
  \bibinfo{author}{\bibfnamefont{R.}~\bibnamefont{Neuhausen}},
  \bibinfo{journal}{Nucl. Phys.} \textbf{\bibinfo{volume}{A327}},
  \bibinfo{pages}{439} (\bibinfo{year}{1979}).

\bibitem[{\citenamefont{Yen et~al.}(1974)\citenamefont{Yen, Cardman, Kalinsky,
  Legg, and Bockelman}}]{Ye74}
\bibinfo{author}{\bibfnamefont{R.}~\bibnamefont{Yen}},
  \bibinfo{author}{\bibfnamefont{L.~S.} \bibnamefont{Cardman}},
  \bibinfo{author}{\bibfnamefont{D.}~\bibnamefont{Kalinsky}},
  \bibinfo{author}{\bibfnamefont{J.~R.} \bibnamefont{Legg}}, \bibnamefont{and}
  \bibinfo{author}{\bibfnamefont{C.~K.} \bibnamefont{Bockelman}},
  \bibinfo{journal}{Nucl. Phys.} \textbf{\bibinfo{volume}{A235}},
  \bibinfo{pages}{135} (\bibinfo{year}{1974}).

\bibitem[{\citenamefont{Bergstrom and Tomusiak}(1976)}]{Be76}
\bibinfo{author}{\bibfnamefont{J.~C.} \bibnamefont{Bergstrom}}
  \bibnamefont{and} \bibinfo{author}{\bibfnamefont{E.~L.}
  \bibnamefont{Tomusiak}}, \bibinfo{journal}{Nucl. Phys.}
  \textbf{\bibinfo{volume}{A262}}, \bibinfo{pages}{196} (\bibinfo{year}{1976}).

\bibitem[{\citenamefont{Hutcheon and Caplan}(1969)}]{Hu69}
\bibinfo{author}{\bibfnamefont{R.~M.} \bibnamefont{Hutcheon}} \bibnamefont{and}
  \bibinfo{author}{\bibfnamefont{H.~S.} \bibnamefont{Caplan}},
  \bibinfo{journal}{Nucl. Phys.} \textbf{\bibinfo{volume}{A127}},
  \bibinfo{pages}{417} (\bibinfo{year}{1969}).

\bibitem[{\citenamefont{Brown et~al.}(1983)\citenamefont{Brown, Radhi, and
  Wildenthal}}]{Br83}
\bibinfo{author}{\bibfnamefont{B.~A.} \bibnamefont{Brown}},
  \bibinfo{author}{\bibfnamefont{R.}~\bibnamefont{Radhi}}, \bibnamefont{and}
  \bibinfo{author}{\bibfnamefont{B.~H.} \bibnamefont{Wildenthal}},
  \bibinfo{journal}{Phys. Rep.} \textbf{\bibinfo{volume}{101}},
  \bibinfo{pages}{313} (\bibinfo{year}{1983}).

\bibitem[{\citenamefont{Amos and Berge}(1983)}]{Am83}
\bibinfo{author}{\bibfnamefont{K.}~\bibnamefont{Amos}} \bibnamefont{and}
  \bibinfo{author}{\bibfnamefont{L.}~\bibnamefont{Berge}},
  \bibinfo{journal}{Phys. Lett.} \textbf{\bibinfo{volume}{127B}},
  \bibinfo{pages}{299} (\bibinfo{year}{1983}).

\bibitem[{\citenamefont{Chien and Khoa}(2009)}]{Ch09}
\bibinfo{author}{\bibfnamefont{N.~D.} \bibnamefont{Chien}} \bibnamefont{and}
  \bibinfo{author}{\bibfnamefont{D.~T.} \bibnamefont{Khoa}},
  \bibinfo{journal}{Phys. Rev. C} \textbf{\bibinfo{volume}{79}},
  \bibinfo{pages}{034314} (\bibinfo{year}{2009}).

\bibitem[{\citenamefont{Bernstein et~al.}(1981)\citenamefont{Bernstein, Brown,
  and Madsen}}]{Be81}
\bibinfo{author}{\bibfnamefont{A.~M.} \bibnamefont{Bernstein}},
  \bibinfo{author}{\bibfnamefont{V.~R.} \bibnamefont{Brown}}, \bibnamefont{and}
  \bibinfo{author}{\bibfnamefont{V.~A.} \bibnamefont{Madsen}},
  \bibinfo{journal}{Comments Nucl. Part. Phys.} \textbf{\bibinfo{volume}{11}},
  \bibinfo{pages}{203} (\bibinfo{year}{1981}).

\bibitem[{\citenamefont{Marechal et~al.}(1999)}]{Ma99}
\bibinfo{author}{\bibfnamefont{F.}~\bibnamefont{Marechal}}
  \bibnamefont{et~al.}, \bibinfo{journal}{Phys. Rev. C}
  \textbf{\bibinfo{volume}{60}}, \bibinfo{pages}{034615}
  (\bibinfo{year}{1999}).

\bibitem[{\citenamefont{Iwasa et~al.}(2008)}]{Iw08}
\bibinfo{author}{\bibfnamefont{N.}~\bibnamefont{Iwasa}} \bibnamefont{et~al.},
  \bibinfo{journal}{Phys. Rev. C} \textbf{\bibinfo{volume}{78}},
  \bibinfo{pages}{024306} (\bibinfo{year}{2008}).

\bibitem[{\citenamefont{Suda and Wakasugi}(2005)}]{Su05}
\bibinfo{author}{\bibfnamefont{T.}~\bibnamefont{Suda}} \bibnamefont{and}
  \bibinfo{author}{\bibfnamefont{M.}~\bibnamefont{Wakasugi}},
  \bibinfo{journal}{Prog. Part. Nucl. Phys.} \textbf{\bibinfo{volume}{55}},
  \bibinfo{pages}{417} (\bibinfo{year}{2005}).

\bibitem[{\citenamefont{Karataglidis et~al.}(2007)\citenamefont{Karataglidis,
  Kim, and Amos}}]{Ka07}
\bibinfo{author}{\bibfnamefont{S.}~\bibnamefont{Karataglidis}},
  \bibinfo{author}{\bibfnamefont{Y.~J.} \bibnamefont{Kim}}, \bibnamefont{and}
  \bibinfo{author}{\bibfnamefont{K.}~\bibnamefont{Amos}},
  \bibinfo{journal}{Nucl. Phys.} \textbf{\bibinfo{volume}{A 793}},
  \bibinfo{pages}{40} (\bibinfo{year}{2007}).

\bibitem[{\citenamefont{Amos et~al.}(2007)\citenamefont{Amos, Faessler, and
  Rodin}}]{Am07}
\bibinfo{author}{\bibfnamefont{K.}~\bibnamefont{Amos}},
  \bibinfo{author}{\bibfnamefont{A.}~\bibnamefont{Faessler}}, \bibnamefont{and}
  \bibinfo{author}{\bibfnamefont{V.}~\bibnamefont{Rodin}},
  \bibinfo{journal}{Phys. Rev. C} \textbf{\bibinfo{volume}{76}},
  \bibinfo{pages}{014604} (\bibinfo{year}{2007}).

\end{thebibliography}

\end{document}